\documentclass[twoside,twocolumn,9pt]{article}

\usepackage{amssymb}
\usepackage{graphicx}
\usepackage{wrapfig}
\usepackage{amsmath}
\usepackage{blindtext}
\usepackage{comment}
\usepackage{ulem}
\usepackage{amsfonts}
\usepackage{mathtools}
\usepackage{booktabs}
\usepackage{multirow}
\usepackage{subcaption}
\usepackage{indentfirst}

\usepackage{extsizes}
\usepackage[super,sort&compress,comma]{natbib} 
\usepackage[version=3]{mhchem}
\usepackage[left=1.5cm, right=1.5cm, top=1.785cm, bottom=2.0cm]{geometry}
\usepackage{balance}
\usepackage{mathptmx}
\usepackage{sectsty}
\usepackage{graphicx} 
\usepackage{lastpage}
\usepackage[format=plain,justification=justified,singlelinecheck=false,font={stretch=1.125,small,sf},labelfont=bf,labelsep=space]{caption}
\usepackage{float}
\usepackage{fancyhdr}
\usepackage{fnpos}
\usepackage[english]{babel}
\addto{\captionsenglish}{%
  \renewcommand{\refname}{Notes and references}
}
\usepackage{array}
\usepackage{droidsans}
\usepackage{charter}
\usepackage[T1]{fontenc}
\usepackage[usenames,dvipsnames]{xcolor}
\usepackage{setspace}
\usepackage[compact]{titlesec}
\usepackage{hyperref}
\usepackage[nolist,nohyperlinks]{acronym}
\usepackage[acronym]{glossaries}
\makeglossaries

\usepackage{epstopdf}

\definecolor{cream}{RGB}{222,217,201}

\begin{document}

\newacronym{ml}{ML}{Machine Learning}
\newacronym{dl}{DL}{Deep Learning}
\newacronym{dt}{DT}{Decision Tree}
\newacronym{rf}{RF}{Random Forest}
\newacronym{gbdt}{GBDT}{Gradient Boosting Decision Tree}
\newacronym{xgb}{XGB}{XGBoost}
\newacronym{tp}{TP}{True Positive}
\newacronym{tn}{TN}{True Negative}
\newacronym{fp}{FP}{False Positive}
\newacronym{fn}{FN}{False Negative}
\newacronym{shap}{SHAP}{Shapley Addictive Explanations}
\newacronym{ae}{AE}{Autoencoder}
\newacronym{vae}{VAE}{Variational Autoencoder}
\newacronym{cvae}{CVAE}{Conditional Variational Autoencoder}
\newacronym{dscvae}{DSCVAE}{Double Space Conditional Variational Autoencoder}
\newacronym{mse}{MSE}{Mean Square Error}
\newacronym{kld}{KLD}{Kullback-Leibler Divergence}
\newacronym{ce}{CE}{Cross Entropy}
\newacronym{ir}{IR}{Imbalanced Ratio}
\newacronym{ann}{ANN}{Artificial Neural Network}
\newacronym{gan}{GAN}{Generative adversarial network}
\newcommand{\rev}{\textcolor{blue}}
\pagestyle{fancy}
\thispagestyle{plain}
\fancypagestyle{plain}{
\renewcommand{\headrulewidth}{0pt}
}

\makeFNbottom
\makeatletter
\renewcommand\LARGE{\@setfontsize\LARGE{15pt}{17}}
\renewcommand\Large{\@setfontsize\Large{12pt}{14}}
\renewcommand\large{\@setfontsize\large{10pt}{12}}
\renewcommand\footnotesize{\@setfontsize\footnotesize{7pt}{10}}
\makeatother

\renewcommand{\thefootnote}{\fnsymbol{footnote}}
\renewcommand\footnoterule{\vspace*{1pt}%
\color{cream}\hrule width 3.5in height 0.4pt \color{black}\vspace*{5pt}} 
\setcounter{secnumdepth}{5}

\makeatletter 
\renewcommand\@biblabel[1]{#1}            
\renewcommand\@makefntext[1]%
{\noindent\makebox[0pt][r]{\@thefnmark\,}#1}
\makeatother 
\renewcommand{\figurename}{\small{Fig.}~}
\sectionfont{\sffamily\Large}
\subsectionfont{\normalsize}
\subsubsectionfont{\bf}
\setstretch{1.125} 
\setlength{\skip\footins}{0.8cm}
\setlength{\footnotesep}{0.25cm}
\setlength{\jot}{10pt}
\titlespacing*{\section}{0pt}{4pt}{4pt}
\titlespacing*{\subsection}{0pt}{15pt}{1pt}

\fancyfoot{}
\fancyfoot[LO,RE]{\vspace{-7.1pt}\includegraphics[height=9pt]{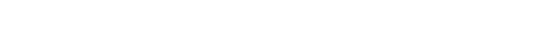}}
\fancyfoot[CO]{\vspace{-7.1pt}\hspace{11.9cm}\includegraphics{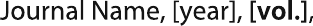}}
\fancyfoot[CE]{\vspace{-7.2pt}\hspace{-13.2cm}\includegraphics{head_foot/RF}}
\fancyfoot[RO]{\footnotesize{\sffamily{1--\pageref{LastPage} ~\textbar  \hspace{2pt}\thepage}}}
\fancyfoot[LE]{\footnotesize{\sffamily{\thepage~\textbar\hspace{4.65cm} 1--\pageref{LastPage}}}}
\fancyhead{}
\renewcommand{\headrulewidth}{0pt} 
\renewcommand{\footrulewidth}{0pt}
\setlength{\arrayrulewidth}{1pt}
\setlength{\columnsep}{6.5mm}
\setlength\bibsep{1pt}

\makeatletter 
\newlength{\figrulesep} 
\setlength{\figrulesep}{0.5\textfloatsep} 

\newcommand{\topfigrule}{\vspace*{-1pt}%
\noindent{\color{cream}\rule[-\figrulesep]{\columnwidth}{1.5pt}} }

\newcommand{\botfigrule}{\vspace*{-2pt}%
\noindent{\color{cream}\rule[\figrulesep]{\columnwidth}{1.5pt}} }

\newcommand{\dblfigrule}{\vspace*{-1pt}%
\noindent{\color{cream}\rule[-\figrulesep]{\textwidth}{1.5pt}} }

\makeatother

\twocolumn[
  \begin{@twocolumnfalse}
{\includegraphics[height=30pt]{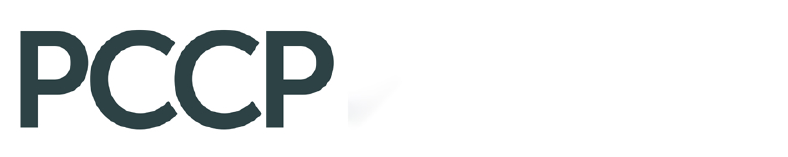}\hfill\raisebox{0pt}[0pt][0pt]{\includegraphics[height=55pt]{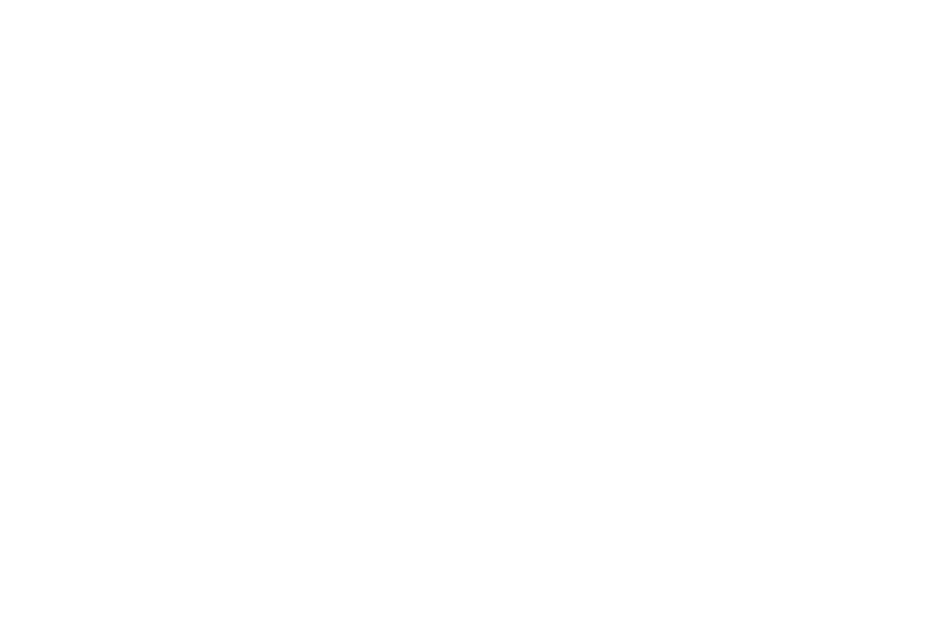}}\\[1ex]
\includegraphics[width=18.5cm]{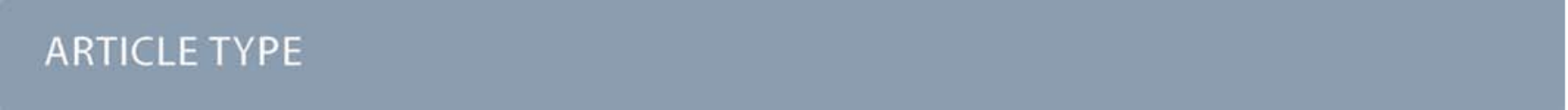}}\par
\vspace{1em}
\sffamily
\begin{tabular}{m{4.5cm} p{13.5cm} }

\includegraphics{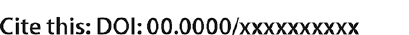} & \noindent\LARGE{\textbf{Analyzing drop coalescence in microfluidic device with a deep learning generative model}} \\
\vspace{0.3cm} & \vspace{0.3cm} \\

 & \noindent\large{Kewei Zhu\textit{$^{a}$}, Sibo Cheng$^{\ast}$\textit{$^{b}$}, Nina Kovalchuk\textit{$^{c}$},  Mark Simmons\textit{$^{c}$},  Yi-Ke Guo\textit{$^{b}$}, Omar K. Matar\textit{$^{d}$}, Rossella Arcucci\textit{$^{e}$}
 }\\

\includegraphics{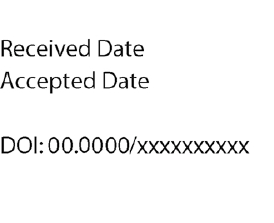} & \noindent\normalsize{Predicting drop coalescence based on process parameters is crucial for experiment design in chemical engineering. However, predictive models can suffer from the lack of training data and more importantly, the label imbalance problem. In this study, we propose the use of deep learning generative models to tackle this bottleneck by training the predictive models using generated synthetic data. A novel generative model, named double space conditional variational autoencoder (DSCVAE) is developed for labelled tabular data. By introducing label constraints in both the latent and the original space, DSCVAE is capable of generating consistent and realistic samples compared to standard conditional variational autoencoder (CVAE). Two predictive models, namely random forest and gradient boosting classifiers, are enhanced on synthetic data and their performances are evaluated on real experimental data. Numerical results show that considerable improvement in prediction accuracy can be achieved by using synthetic data and the proposed DSCVAE clearly outperforms the standard CVAE. This research clearly brings more insight into handling imbalanced data for classification problems, especially in chemical engineering.} \\

\end{tabular}

 \end{@twocolumnfalse} \vspace{0.6cm}
 ]

\renewcommand*\rmdefault{bch}\normalfont\upshape
\rmfamily
\section*{}
\vspace{-1cm}

\begin{NoHyper}
\footnotetext{\textit{$^{a}$~Department of Computer Science, University of York, UK}}
\footnotetext{\textit{$^{b}$~Data Science Institute, Department of Computing, Imperial College London, UK}}
\footnotetext{\textit{$^{c}$~School of Chemical Engineering, University of Birmingham, UK}}
\footnotetext{\textit{$^{d}$~Department of Chemical Engineering Imperial College London, UK}}
\footnotetext{\textit{$^{e}$~Department of Earth Science \& Engineering, Imperial College London, UK}}
\footnotetext{\textit{$^*$~Corresponding: sibo.cheng@imperial.ac.uk}}
\end{NoHyper}



\section{Introduction}

Drop coalescence is of high industrial importance as it determines emulsion stability. Extensive effort has been devoted to comprehending and predicting coalescence under diverse hydrodynamic conditions, with many investigations seeking to minimize or maximize the likelihood of coalescence based on the specific application. Typical areas of industrial interest where coalescence should be avoided are the prediction and maximization of foam/emulsion shelf life (mild hydrodynamic conditions) or product stability during transportation (more challenging hydrodynamic conditions, including, for example, shaking). On the other hand, coalescence is desirable in separation processes, such as the elimination of aqueous droplets that contaminate an oil. 

Microfluidics provide a unique opportunity to study drop coalescence under well-controlled hydrodynamic conditions and if necessary to account for the fate of each formed doublet. The use of microfluidic platforms allows for the exploration of thousands of coalescence instances with minimal material usage and energy consumption. Drop coalescence is used in microfluidics for triggering/quenching chemical reactions within a drop or for cell screening. Various designs of microfluidic chambers are used depending on the aim of investigation ~\cite{krebs2012microfluidic, liu2016droplet, shenoy2016stokes, niu2008pillar}. 

Considering that \acrfull*{ml} technique has widely succeeded in chemistry and chemical engineering~\cite{keith2021combining} with applications spanning from quantum chemistry research~\cite{dral2020quantum} to molecule reaction kinetics~\cite{meuwly2021machine}, we believe that effective ML model also can predict the probability of drop coalescence in the microfluidic device, enabling a considerable reduction of time and cost spent on optimization of microfluidic designs including the cost of energy as well as expensive and hazardous materials used in photo-lithography.

For microfluidic drop reactions or cell screening relying on coalescence, it is obligatory that the rate of drop coalescence is close to 100\%. In such scenarios, the composition of coalescing drops is set as a priority, and there are usually very limited possibilities for adjusting the properties of the continuous phase. Therefore, accurate prediction of drop coalescence based on geometrical and hydrodynamic conditions is pivotal for effective experimental design. In this study, the experimental dataset comprises outcomes of either "coalescence" or "non-coalescence" when two drops interact in a microfluidic coalescence chamber, contingent on process parameters, such as drop size and flow rate.

Here, \acrshort*{ml} can be used for the optimization of parameters of the coalescence chamber and geometry of microfluidic devices in general, as well as flow characteristics maximizing the drop coalescence probability. By construction, different \acrshort*{ml} methods are able to deal with various types of information in chemistry, such as images for phenomena, videos for processes, texts for descriptions, and tabular data for numerical records. For example, Lasso regression and \acrfull*{rf} have been employed to predict drop coalescence based on experimental data~\cite{wikramanayake2020statistical}.  Moreover, the usage of neural networks has helped parameterize the collision merging process~\cite{rodriguez2022parameterization} and surrogate models based on \acrfull*{dl} models are developed to predict the dynamics of drop interactions based on recorded videos of experiments~\cite{zhuang2022ensemble}. Using effective \acrshort*{ml} protocols providing an accurate prediction of coalescence in combination with microfluidics enables in this case considerable reduction in material and energy consumption during formulation optimization. 

Despite the success in a large range of applications, it is widely noticed that imbalanced training data can lead to poor prediction/classification performance in \acrshort*{ml}~\cite{he2009learning}. Examples can be found  in the assessment of chemical-disease relation~\cite{mitra2020multi} and in the classification of drug-induced injury\cite{thakkar2018liver}. To tackle the data imbalance problem, multiple advanced tree-based algorithms~\cite{su2015improving,blaszczynski2015neighbourhood,bader2018biased,ferreira2019adaptive} are developed. These approaches mainly consist of building sub-models trained by partial data to alleviate the risk of overfitting caused by data imbalance. However, in the case of extreme data imbalance, the size of the training datasets in each sub-model is limited due to the constraint of label balance~\cite{batista2004study}. 

Generative models are developed as a method of data augmentation~\cite{wei2019eda}. From this extension, they are employed to address the issue of data imbalance. \acrfull*{gan}~\cite{goodfellow2020generative} is a common method to generate artificial data, of which the generator is used to generate data, and then the discriminator judges if the synthetic data are similar to the real. Although \acrshort*{gan}s reach obvious improvement on imbalance data issue~\cite{moon2020conditional}, the implicit posterior distribution is difficult to optimize and requires large-scale data to converge~\cite{burks2019data}. Another method to create synthetic data is \acrfull*{vae}~\cite{kingma2014auto}. It is proposed with an explicit posterior distribution, whereas VAE cannot generate data with class-level conditions. To enable VAEs to generate class-specific data, \acrfull*{cvae}~\cite{sohn2015learning} is developed with an extra classifier in the original space. Thanks to the constraint, \acrshort*{cvae} can learn conditional representations explicitly in the original space and implicitly in the latent space simultaneously. Benefiting from the conditional representation, \acrshort*{cvae} can generate synthetic data separately and substantially to decrease the impact of imbalanced data issue~\cite{huang2022ada,chen2021trajvae,yang2019improving}. However, the latent space of \acrshort*{cvae} still lacks conditional constraint. Therefore, it is unable to explicitly learn latent conditional representation, which pushes more information stored in the decoder rather than the latent space~\cite{he2022masked}. The recent work of Chagot et al. ~\cite{chagot2022surfactant} generates synthetic data to mitigate the issue of imbalanced data but the challenge associated with latent conditional representation remains. 

To improve the performance of generative models, some studies have begun to focus on latent spaces~\cite{gelada2019deepmdp, bojanowski2018optimizing}. GET-3D~\cite{gao2022get3d} in 3D modeling area does not constrain the outputs only. It is proposed to add an extra constraint in latent space to improve the consistency of outputs. However, all outputs from GET-3D are not according to specific labels due to the target task. For example, GET-3D is capable of generating vivid 3D cars, but there is no label-specific constraint involved. GET-3D just guides the generator to generate similar data without focusing on their class differences. 

In this work, we aim to handle an experiment-based classification problem using \acrshort*{ml} methods. This study is conducted based on tabular data, which have a relatively small sample size and the results (i.e., "coalescence" or "non-coalescence") correspond to a very similar distribution of conditional values. By predicting drop coalescence in the microfluidic device, we explore a new implementation to solve the dilemma of tabular data imbalance. To accomplish the prediction tasks, we use \acrshort*{ml} methods of \acrshort*{rf} and \acrlong*{xgb} for their interpretability and strengths in processing small sample-sized datasets. To address the data imbalance for better predictive performances, we use generative models based on variational autoencoder (VAE) to get more synthetic data. In this stage, we propose a \acrfull*{dscvae} consisting of a standard VAE and two explicit constraints on both the latent space and the original space. Its novel latent constraint guides the latent space to learn conditional representation and enables the latent space of \acrshort*{dscvae} to become more informative. After generating more training data using \acrshort*{dscvae}, the accuracy of the predictive models improved by 7.5\% and 8.5\%, respectively. Through sensitivity analysis based on \acrfull*{shap} values, the contribution of these synthetic data to predictive models is more consistent with microfluidics experiments.
\begin{figure*}[!htb]
    \centering
    \subfloat[]{\includegraphics[width=2.5 in]{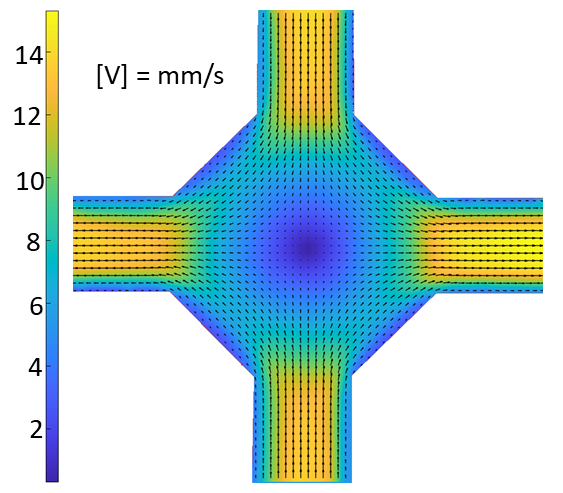} \label{fig:e_a}}
    \subfloat[]{\includegraphics[width=2.5 in]{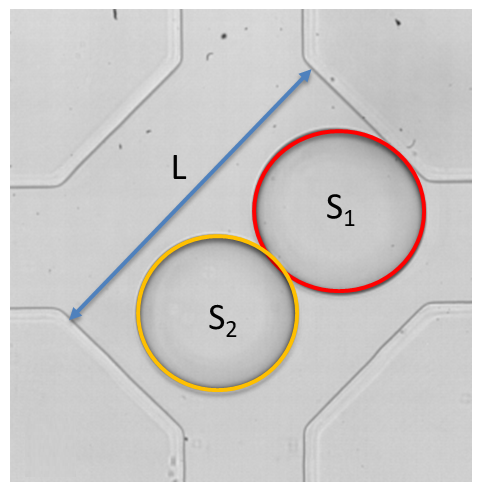} \label{fig:e_b}}
    \caption{Drop coalescence experiments in microfluidic device}
    \label{fig:experiment}
\end{figure*}

The rest of the paper is organized as follows. 
Section~\ref{Experiments and Dataset} describes the experimental process and the initial data that originated from the experiment. Section~\ref{mtd:method} introduces the \acrlong*{ml} methods involved in this task, including two tree-based predictive algorithms for predicting coalescence results and our novel \acrshort*{dscvae} model for synthetic data. Section~\ref{NumericalResults} exhibits the results of our numerical experiments. Starting with implementation details, we trace the training process of generative models, discuss the performance of predictors, and analyze the impact of the initial and generative datasets on the predictors via SHAP.

\section{Experiments and Dataset}\label{Experiments and Dataset}
\subsection{Experiments}\label{Experiments}

Coalescence experiments were carried out in a microfluidic device with rectangular channels made of PDMS using standard soft lithography~\cite{kim2008soft}. Aqueous drops (double distilled water from water still Aquatron A 4000 D, Stuart)  in a continuous phase of silicone oil (5 cSt, Aldrich) were formed using flow-focusing~\cite{kovalchuk2019drop} in two symmetrical cross-junctions and met within a coalescence chamber shown in Fig.\ref{fig:experiment}. Drop movement and coalescence were recorded using a high-speed video camera Photron SA-5 connected to an inverted microscope Nikon Eclipse Ti2-U at 1000 frames per second and spatial resolution of 2 µm/pixel. The images were processed in ImageJ~\cite{schneider2012nih} to find drop sizes. 

The chamber has two entrances or input channels (top and bottom channel in Fig.\ref{fig:experiment}) and two exits or output channels (left and right channel in Fig.\ref{fig:experiment}) of width 315 ± 5 µm and depth 140 ± 5 µm. The chamber size, $\mathbf{L}$, shown in Fig.\ref{fig:e_b} was in the range of 901 ± 11 µm. The flow field in the chamber and adjacent channels is shown in Fig. 1a,  with a velocity magnitude shown in mm/s. It was measured using Ghost Particle Velocimetry, the non-invasive technique using as a tracer the speckle pattern produced by light scattered by particles smaller than the diffraction limit~\cite{buzzaccaro2013ghost,kovalchuk2018study}. As Fig.\ref{fig:e_a} shows, there is a stagnation point in the centre of the chamber due to flow symmetry. Drops arrive into the chamber with various time delays between them caused by fluctuations in the flow rate of continuous and dispersed phases supplied by syringe pumps (Al-4000, World Precision Instruments), inevitable deviations in channel size, and fluctuations in channel resistance due to drop presence. 

In an ideal situation, the drops move along the symmetry axis of input channels and the first drop arriving in the chamber is trapped at the stagnation point till the arrival of the second drop. In reality, the flow fluctuations and effect of the second drop on the flow field in the chamber result in drops encounter at various positions around the stagnation point. Following an encounter, the drops form a doublet, i.e., start to move together. If the doublet axis coincides with that of the input channels, which is also one of the chamber symmetry axes, the doublet is trapped in the compression flow provided by the continuous phase. In this case, the coalescence probability is 100\% and compositions of continuous and dispersed phases affect only the time span between the doublet formation and coalescence. Because of ever-present fluctuations as well as the inevitable small asymmetries of real devices, the doublet begins to rotate/translate to one of the output channels; a doublet rotated and moved from its initial position is shown in Fig.\ref{fig:e_b}. Doublet rotation results in its transfer from a region of compression flow to one of extensional flow when its axis coincides with the axis of output channels. The extensional flow leads to drop detachment resulting in a non-coalescence event. 

In general, the outcome of the drop encounter depends on the relative time scales of drainage of the continuous phase from the film separating the drops and the transition of the doublet from the region of compression flow to that of extensional flow: if the doublet does not rotate at all, the coalescence probability is 100\%; if it rotates very fast, the coalescence probability is zero, i.e., the coalescence probability increases with decreasing flow rate. However, the very slow flows are incompatible with  the  requirement of high throughput being a measure of the efficiency of microfluidic devices. Moreover, in a certain range of flow rates, the drainage time decreases with an increase of continuous phase flow rate due to inertial effects~\cite{yi2020efficient} diminishing, or possibly even reversing, the dependence of the coalescence probability on the flow rate. 



Therefore, process optimization is necessary to determine the conditions under which coalescence probability remains large while maintaining a sufficiently high throughput. The optimization parameters include $\mathbf{x}_\textrm{flow}$, the total flow rate; $\mathbf{x}_\textrm{drop1}$ and $\mathbf{x}_\textrm{drop2}$, the drop diameters; and $\mathbf{x}_\textrm{dt}$, the time delay between drops. The total flow rate is calculated as the sum of the flow rates of continuous and dispersed phases, with a syringe pump dispensing accuracy of ±1\%. The drop sizes are determined from their area in the plane of observation, $\mathbf{S1}$ and $\mathbf{S2}$ in Fig.\ref{fig:e_b}, with a maximum error of ±1\%. The drop diameters are normalized by the chamber width. The time delay, $\mathbf{x}_\textrm{dt}$ between drops has an uncertainty of 2 ms.


\subsection{Dataset} \label{Dataset}
The experimental dataset consists of 1531 samples in total, including one label $\textbf{y}$, influenced by four features $\textbf{X}^{\textrm{features}}=[\mathbf{x}_\textrm{flow}, \mathbf{x}_\textrm{drop1}, \mathbf{x}_\textrm{drop2}, \mathbf{x}_\textrm{dt}]$ in various ranges. The label $\textbf{y}$ has two classes:  "coalescence" and "non-coalescence". The features are scaled for their comparability by min-max normalization according to
\begin{equation}
\mathbf{x}_\textrm{scaled}=\frac{\mathbf{x}-\mathbf{x}_{\min}}{\mathbf{x}_{\max}-\mathbf{x}_{\min}}, \text{ where }\mathbf{x}\in {\{\mathbf{x}_\textrm{flow}, \mathbf{x}_\textrm{drop1}, \mathbf{x}_\textrm{drop2}, \mathbf{x}_\textrm{dt}\}}
\label{eq:normal}
\end{equation}
here $\mathbf{x}_{\max}$ and $\mathbf{x}_{\min}$ for its maximum value and minimum value of $\mathbf{x}$, and $\mathbf{x}_{scaled}$ for its normalised outcomes. 
\begin{figure}[!htb]
    \centering
    \subfloat[Sample proportion]{\includegraphics[width=1.6 in]{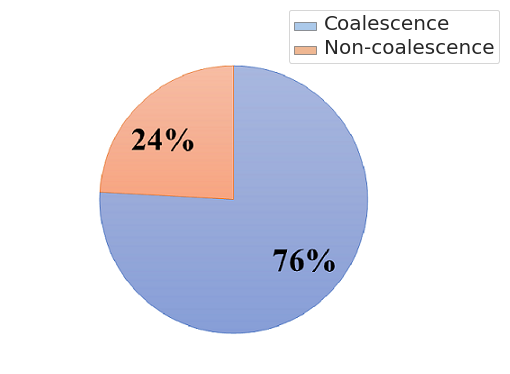} \label{fig:Pie}}
    \subfloat[Frequency distribution]{\includegraphics[width=1.6 in]{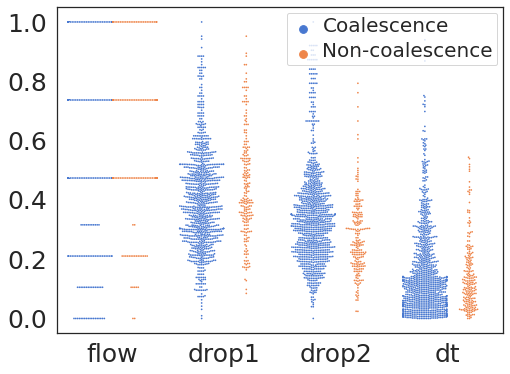} \label{fig:Swarm}}\\
    \subfloat[Median and quartiles]{\includegraphics[width=1.6 in]{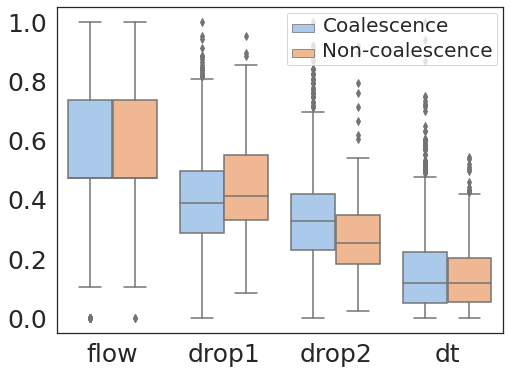} \label{fig:Box}}
    \subfloat[Relative frequency distribution]{\includegraphics[width=1.6 in]{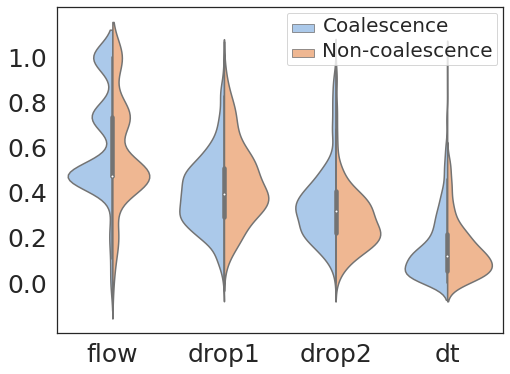} \label{Fig.Violin}}
    \caption{Experimental dataset characteristics}
    \label{fig:dataset}
\end{figure}

The statistical characteristics of binary classifications are given in detail in Fig.\ref{fig:dataset}. Figure \ref{fig:Pie} shows the imbalance percentages of "coalescence" and "non-coalescence" samples, where "coalescence" accounts for 76\% and "non-coalescence" only accounts for 24\%. The value distributions of four features compared between two labels are shown in Fig.\ref{fig:Swarm}. Fig.\ref{fig:Box} shows the median, quartiles and whiskers of $\textbf{X}^{\textrm{features}}$. These values of the two types of labels under each feature are not significantly different. Fig.\ref{Fig.Violin} shows $\textbf{X}^{\textrm{features}}$ are not subject to the standard normal distribution, and confirms again that the distributions are similar with different labels. 
To sum up, $\textbf{X}^{\textrm{features}}$ is imbalanced in volume regarding two opposite classes, though identical in distributions. In other words, there are significantly more samples of "coalescence" compare to "non-coalescence" in the training dataset, and it is difficult to separate the two classes of data by simply looking at their input distributions. The latter makes the prediction task even more challenging.

\acrshort*{ml} models must be evaluated using data that have not been used for training. Moreover, a validation set is needed for tuning the hyperparameters of the predictive models. As there is no dominant experimental result in actuality, the dataset is split into a balanced test set and a balanced validation set ("balanced" here means with the same number of "coalescence" and "non-coalescence"), instead of being divided proportionally. The remaining samples are used as the training set. The \acrfull*{ir} here is equal to the number of major results ("coalescence") divided by the number of minor results ("non-coalescence"). Thus, the whole dataset is split into three mutually-independent sets, namely the Balanced training dataset, Validation dataset and Test dataset, for subsequent procedures (shown in table~\ref{tab:dataset}). It is worth mentioning that all comparative models in this paper are trained on this balanced training set to avoid overfitting. 
\begin{table}[!htb]
    \centering
        \caption{Dataset Split}
    \resizebox{0.49\textwidth}{!}{%
    \begin{tabular}{rcccc}\hline
         & \textbf{Coalescence} & \textbf{Non-coalescence} & \textbf{\acrshort*{ir}} & \textbf{Total} \\\hline
        \textbf{Total dataset} & 1162 & 369 & 3.15 & 1531 \\
        Training dataset & 1012 & 219 & 4.62 & 1231 \\
        *Balanced training dataset & 219 & 219 & 1 & 438 \\
        Validation dataset & 50 & 50 & 1 & 100 \\
        Test dataset & 100 & 100 & 1 & 200 \\\hline
    \end{tabular}%
    }
    \label{tab:dataset}
\end{table}

After dataset splitting, \acrshort*{ir}  becomes larger from 3.15 of the total experimental set to 4.62 of the training dataset which makes the classification task more difficult.

\section{Methods} \label{mtd:method}
In this section, we introduce two advanced tree-based models, \acrshort*{rf} and \acrlong*{xgb}, to make predictions for coalescence results. In addition, we design a \acrshort*{vae} variant, \acrshort*{dscvae}, to solve the aforementioned insufficiency and imbalance problem by generating synthetic data. 
\subsection{Predictive models} \label{mtd:predictive}
Tree-based models belong to the class of widely-used algorithms for assessing tabular data, which outperform neural networks for small and intermediate-sized datasets\cite{grinsztajn2022tree}. Therefore, they are applied to process the small-scale data of this study.
The \acrfull*{dt} is a classical \acrshort*{ml} algorithm and has natural interpretability. It has been used in \acrshort*{ml} models for classification and regression~\cite{patel2018study,gong2022efficient}. To avoid overfitting, some parameters, such as max depth $\mathbf{d}_\textrm{max}$, should be set to enhance its robustness\cite{pal2003assessment}. Nevertheless, since minor changes may lead to an entirely disparate tree\cite{turney1995bias}, some variants combined with ensemble learning approaches, such as \acrshort*{rf} (see in section~\ref{mtd:rf}) and \acrlong*{xgb} (see in section~\ref{mtd:xgb}), have been utilized to make predictions. The former and latter adopt "bagging" and  "boosting" algorithms, respectively.

\begin{figure}[htb!]
    \centering
    \subfloat[\acrfull*{rf}]{\includegraphics[height=1.7 in]{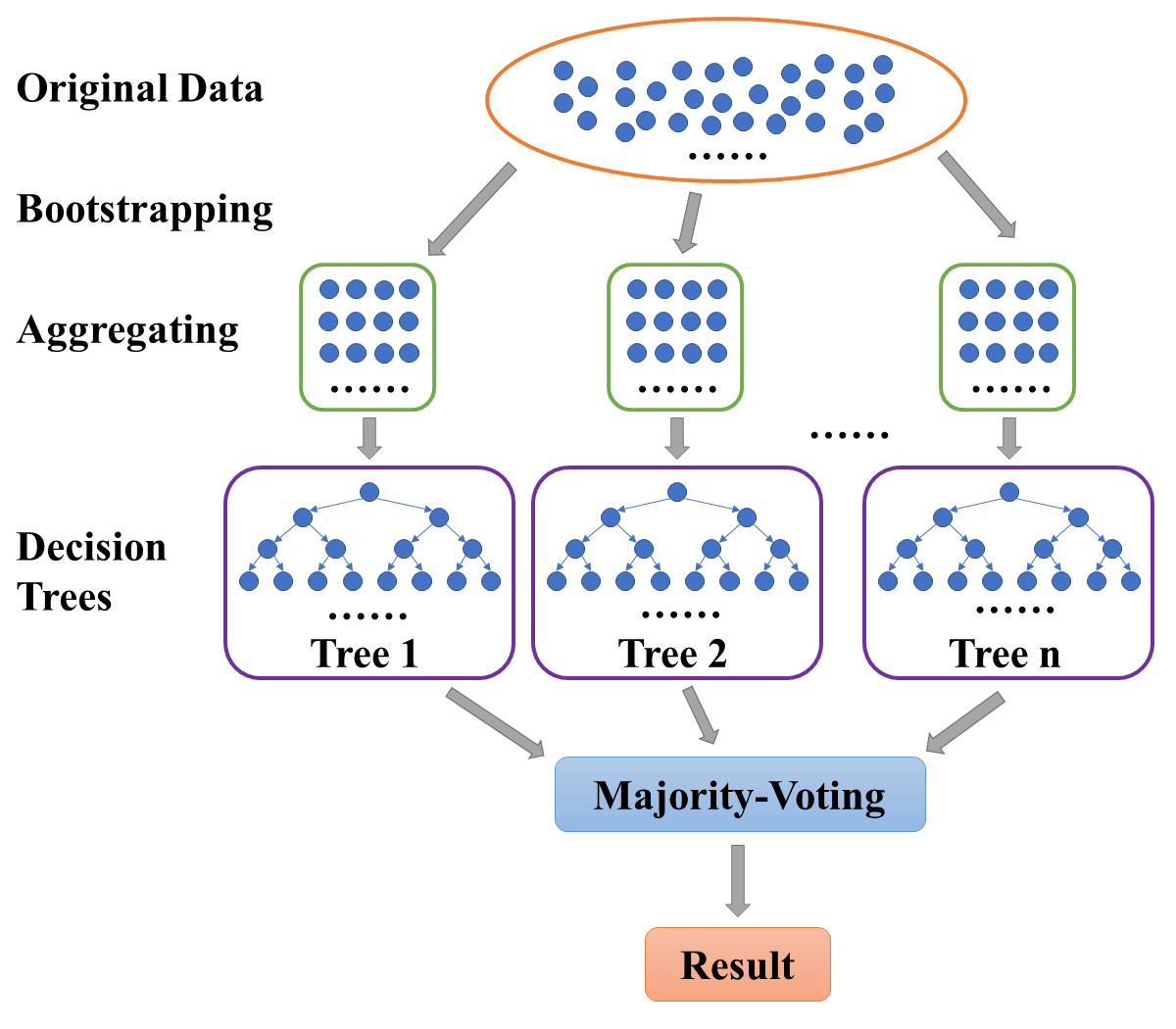} \label{fig:rf}}
    \subfloat[\acrlong*{xgb}]{\includegraphics[height=1.7 in]{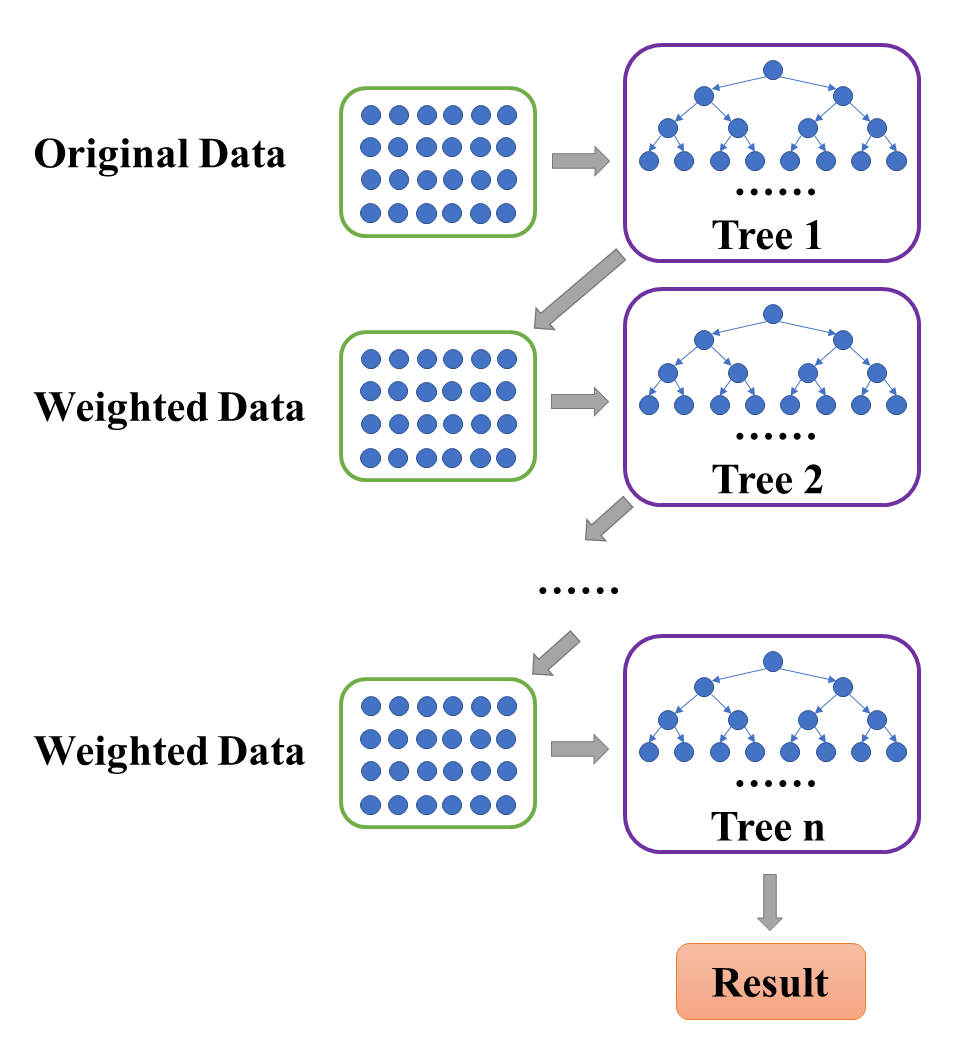} \label{fig:xgb}}
    \caption{Flowchart of the predictive models}
\end{figure}
\subsubsection{Random forest}\label{mtd:rf}
\acrshort*{rf} is a widely applied model focusing on imbalanced data thanks to the "bagging" technique\cite{ho1995random} (Fig.\ref{fig:rf}). Here, the term "Bagging" corresponds to the abbreviation of "Bootstrap aggregating", that is, multiple models trained by bootstrap sampling in a parallel process\cite{breiman1996bagging,cheng2022parameter}. An  \acrshort*{rf} selects an irregular portion of samples from the training set with put-backs and randomly selects some features for training; the number of sub-models is $\mathbf{n}_\textrm{estimators}$. The results of sub-models vary, because of different sub-datasets. After that, a prediction is made by taking a majority vote on classification trees. The benefit of this approach is that it not only settles the over-fitting problems but also helps to reduce prediction errors in general~\cite{breiman2001random}. 
\subsubsection{XGBoost}\label{mtd:xgb}
Unlike "bagging" where submodels expand in parallel, "boosting" is a sequential structure. It has been used in \acrfull*{gbdt} to decrease variance and bias of tree models\cite{friedman2001greedy}. The new tree is built on the basis of the previous one, to decrease the residuals to the target. \acrlong*{xgb} is an efficient implementation of \acrshort*{gbdt} (Fig.\ref{fig:xgb}) based on a gradient boosting algorithm~\cite{chen2015xgboost}. It is also robust to over-fitting. 
 Its important parameters also include $\mathbf{d}_\textrm{max}$ and $\mathbf{n}_\textrm{estimators}$, similar to \acrshort*{rf}. 
\subsubsection{Metrics}
In order to evaluate the prediction performance from different perspectives, and equip data characteristics with interpretability, several common approaches were chosen in this study. 

To introduce model-performance measures for binary classification, we used the following metrics to evaluate the classification performance, including accuracy, precision (Eq.\ref{precision}) from an actual perspective, recall (Eq.\ref{recall}) from a predictive perspective, F1 score (Eq.\ref{f1}), and confusion matrix (Table~\ref{cm}). 
\begin{itemize}
\item \textbf{Confusion matrix} includes four categories. Here, \acrfull*{tp} and \acrfull*{tn} indicate correctly classified labels, and \acrfull*{fp} means the amount of incorrectly predicted positive results, and vice versa, \acrfull*{fn} shows incorrectly classified negative results.
\begin{table}[!htb]
\centering
\caption{An example of confusion Matrix}
\begin{tabular}{c|c|c} 
\hline
Real\textbackslash{}Prediction & Negative & Positive \\ \hline
Negative & \textbf{\acrshort*{tn}} & \textbf{\acrshort*{fp}} \\\hline
Positive & \textbf{\acrshort*{fn}} & \textbf{\acrshort*{tp}} \\\hline
\end{tabular}\
\label{cm}
\end{table}

\item \textbf{Precision} or positive predictive value:
\begin{equation}
\textrm{Precision} =\frac{\mathrm{TP}}{\mathrm{TP}+\mathrm{FP}}
\label{precision}
\end{equation}
\item \textbf{Recall} or true positive rate:
\begin{equation}
\textrm{Recall} =\frac{\mathrm{TP}}{\mathrm{TP}+\mathrm{FN}}
\label{recall}
\end{equation}

\item \textbf{F1 score} is the harmonic mean of precision and recall:
\begin{equation}\label{f1}
\mathrm{F1}_{score} = \frac{2}{\frac{1}{\textrm{Precision}}+\frac{1}{\textrm{Recall}}}
= 2 \times \frac{\mathrm{\textrm{Precision}} \times \mathrm{TPR}}{\mathrm{PPV}+\mathrm{\textrm{Recall}}}
\end{equation}
\end{itemize}
\textbf{SHAP values} (\acrshort*{shap}) is based on the classical Shapley value in game theory~\cite{kuhn1953contributions, lundberg2020local2global}. It is used in this study to interpret feature influences toward predictive results by accounting for how each feature affects the tree model. 
\subsection{Generative model} \label{mtd:generative}
The imbalance (Fig.\ref{fig:Pie},\ref{fig:Swarm}) between the two classes of labels caused a preference in the tree model prediction. Moreover, balancing the data (i.e., deleting the excess) reduces the number of majority samples and thus the total number of samples, which could cause the training dataset insufficient for predictive models. This is the reason that the number of samples should be increased with the generative model. 

Generative models, namely \acrlong*{vae}s, are employed in this study to address data imbalance by generating data with different labels separately. These data are then used as inputs for predictive models to enhance performance. 
\begin{figure}[!htb]
    \begin{flushleft}
    \subfloat[Autoencoder (AE)]{\includegraphics[height=1.2 in]{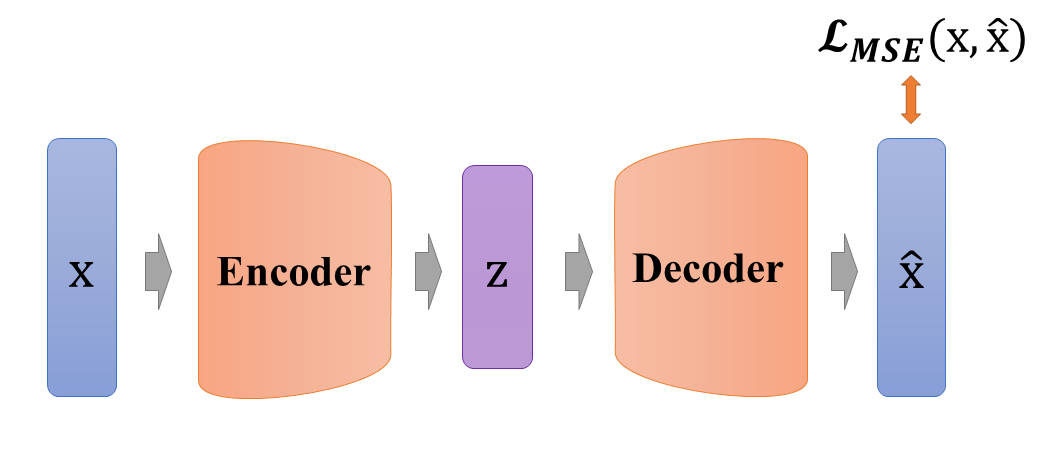} \label{fig:ae}}\\
    \subfloat[Variational Autoencoder (VAE)]{\includegraphics[height=1.2 in]{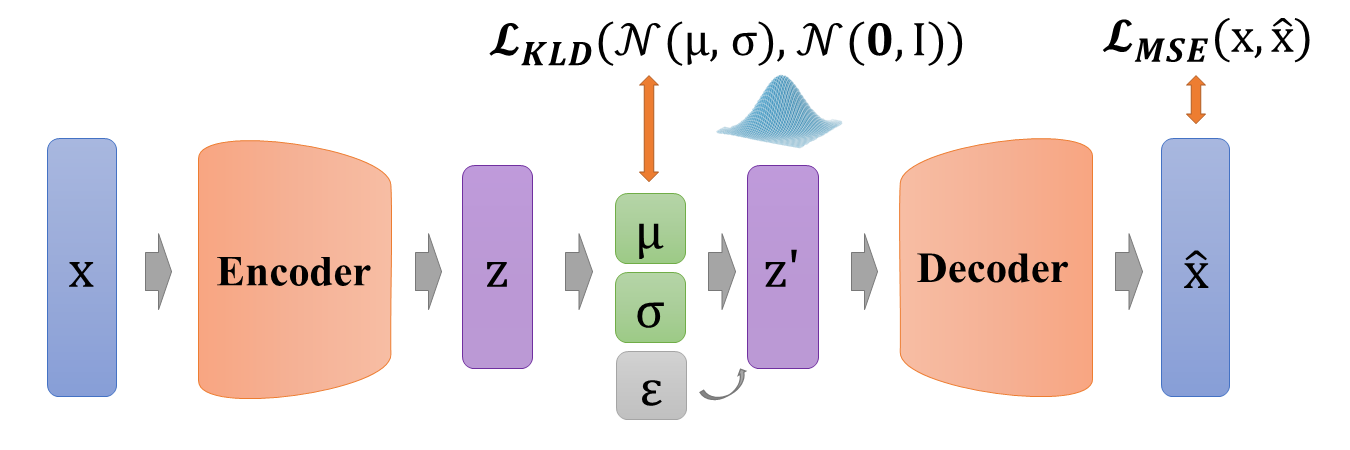} \label{fig:vae}}\\
    \subfloat[Conditional \acrshort*{vae} (\acrshort*{cvae})]{\includegraphics[width=3.6 in]{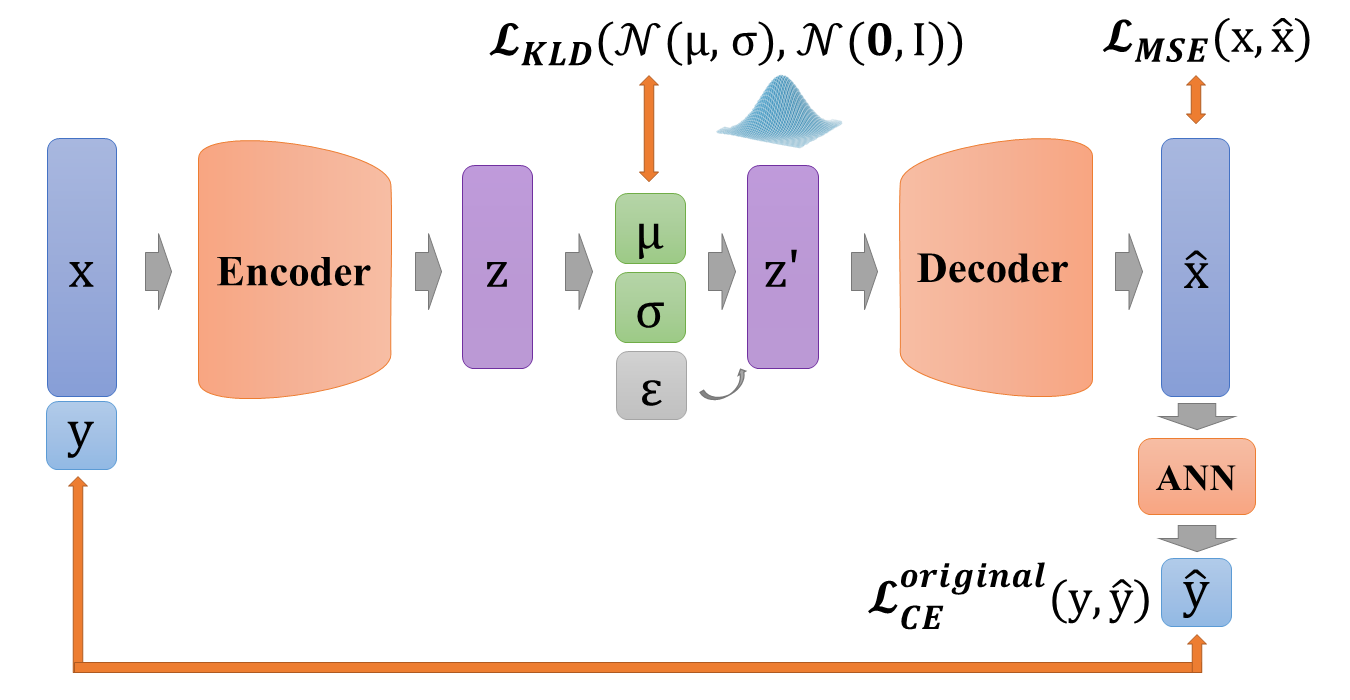} \label{fig:cvae}}
    \caption{Existing \acrshort*{vae} structures}
    \end{flushleft}
\end{figure}
\subsubsection{Variational Autoencoder (VAE)}\label{mtd:vae}
The \acrfull*{ae} is a feed-forward neural network belonging to unsupervised learning\cite{kramer1991nonlinear}. A typical \acrshort*{ae} contains symmetrical architectures as shown in Fig.\ref{fig:ae}. The input $\mathbf{x}$ is encoded to obtain the latent variable $\mathbf{z}$, from which the synthetic output $\hat{\mathbf{x}}$ is decoded. Therefore, an encoder $E$ and a decoder $D$ are represented by
\begin{align} \label{eq:MSE}
    \mathbf{z} = E(\mathbf{x}) \quad \textrm {and} \quad  \hat{\mathbf{x}} = D(\mathbf{z})
\end{align}
This network is trained by minimzing the reconstruction error $L(\mathbf{x}, \hat{\mathbf{x}})$ between  $\hat{\mathbf{x}}$ and $\mathbf{x}$,  usually using \acrfull*{mse}:  
\begin{align}
    \mathcal{L}_{MSE}(\mathbf{x}, \hat{\mathbf{x}})=\frac{1}{N} \sum_{i=0}^N\left(\mathbf{x}-\hat{\mathbf{x}}_i\right)^2
\end{align}
This loss shows the efficiency of data compression strategy. It is used to compare the difference between model output and input. Thus, \acrshort*{ae}s are traditionally used for dimensionality reduction~\cite{cheng2023generalised,cheng2022data} and feature extraction~\cite{yousefi2017autoencoder} instead of data generation. 

\acrshort*{vae} stems from \acrshort*{ae}, and starts to focus on the distribution of $\mathbf{z}$ in latent space (Fig.\ref{fig:vae})~\cite{kingma2014auto}, which equips \acrshort*{vae} with generation capability. Unlike \acrshort*{ae}, \acrshort*{vae} constructs the distribution of $\mathbf{z}$ in the latent space and resamples $\mathbf{z'}$ over the distribution. Thus, the whole process of \acrshort*{vae} can be expressed as
\begin{align} \label{MSE}
    \mathbf{z} = E(\mathbf{x}) \quad \textrm {and} \quad  \hat{\mathbf{x}} = D(\mathbf{z'})
\end{align}

Latent distribution $q(\mathbf{z}|\mathbf{x})$ is trained to be subject to a posterior distribution $p(\mathbf{z|\mathbf{x}})$, usually assumed Gaussian. The approximation between $q(\mathbf{z}|\mathbf{x})$ and $p(\mathbf{z|\mathbf{x}})$ is measured by \acrfull*{kld} loss as 
\begin{equation}
\mathcal{L}_{KLD}(q(\mathbf{z|x}) \| p(\mathbf{z|x})) = - \frac{1}{2} \sum\left(1+\log (\sigma^2)-\mu^2-\sigma^2 \right) 
\end{equation}

\acrshort*{kld} loss is used to regularize the distribution of hidden variables $q(\mathbf{z}|\mathbf{x})$  in the latent space to converge to a Gaussian distribution $p(\mathbf{z|\mathbf{x}})$. Thus, \acrshort*{vae} enables the model with interpretability. 

Resampled latent variables $\mathbf{z'}$ can be characterized by a normal distribution with mean $\mu$ and variance $\sigma$ denoted as $\mathcal{N}(\mu,\sigma^2)$, with $\epsilon$ sampled from a Gaussian distribution as
\begin{align}
    \mathbf{z'} \approx \mu + \epsilon \cdot \sigma \textrm{, where } \epsilon \sim \mathcal{N}(0,\mathbf{I}) 
    \label{eq:reparameterization}
\end{align}
To sum up, the total loss function of \acrshort*{vae}~\cite{esmaeili2019structured} presents as 
\begin{align}
   \mathcal{L} = \mathcal{L}_{MSE}(\mathbf{x}, \hat{\mathbf{x}}) + \mathcal{L}_{KLD}(q(\mathbf{z|x}) \| p(\mathbf{z|x}))
\end{align}

\acrshort*{vae} can be used not only like \acrshort*{ae}s for dimensionality reduction, but also to generate "consistent" data that are different from inputs because of their latent state representation~\cite{chagot2022surfactant}. Here, by 'consistent', we refer to the generated data whose statistical properties mimic those of the original dataset. However, it is also an unsupervised learning method and cannot handle condition-specific samples~\cite{zhao2017learning} (e.g., "coalescence" and "non-coalescence" in this study) because conditional prerequisites are involved  neither in the encoding nor decoding processes. 
\subsubsection{Conditional variational autoencoder (CVAE)}\label{mtd:cvae}
To modulate the latent distribution according to prior conditions, a \acrshort*{cvae}~\cite{sohn2015learning} is devised to generate data with its conditional information. \acrshort*{cvae} concatenates feature inputs $\mathbf{x}$ with auxiliary condition $\mathbf{y}$, thus, it becomes a supervised learning model. 

In addition, in our study, labels $\mathbf{y}$ are not put directly into the encoder with $\textbf{X}^{\textrm{features}}$ together to prevent label leakage during training. In other words, the label $\mathbf{y}$ is not given to the encoder when constructing latent variables. As shown in Fig.\ref{fig:cvae}, firstly, $\textbf{X}^{\textrm{features}}$ are used as inputs for the encoder only. Then, their corresponding labels $\mathbf{y}$ are used as the target of \acrfull*{ann} classifier results $\mathbf{\hat{y}}$, and this \acrshort*{ann} classifier is jointly trained with the entire \acrshort*{vae} network. The difference is measured by predicted labels $\mathbf{\hat{y}}\in(0,1)$ and probability $\mathcal{V}$ via \acrfull*{ce} loss according to
\begin{equation}
\mathcal{L}_{CE}^{original} = -(\mathbf{\hat{y}} \log \mathcal{V}+(1-\mathbf{\hat{y}}) \log (1-\mathcal{V}))
\end{equation}

Thus, the total loss can be described as
\begin{align}
   \mathcal{L} = \mathcal{L}_{MSE}(\mathbf{x}, \hat{\mathbf{x}}) + \mathcal{L}_{KLD}(q(\mathbf{z|x,y}) \| p(\mathbf{z|x,y})) + \mathcal{L}_{CE}^{original}
\end{align}

\acrshort*{cvae} helps generate synthetic data by conditions, and it changes the distribution of the latent space indirectly. However, it is not sufficient to distinguish distributions that are very similar in the original space based on its conditional restrictions. 
\subsubsection{Double space conditional variational autoencoder (DSCVAE)}\label{mtd:dscvae}
\begin{figure}[t]
  \centering
  \includegraphics[width=3.6 in]{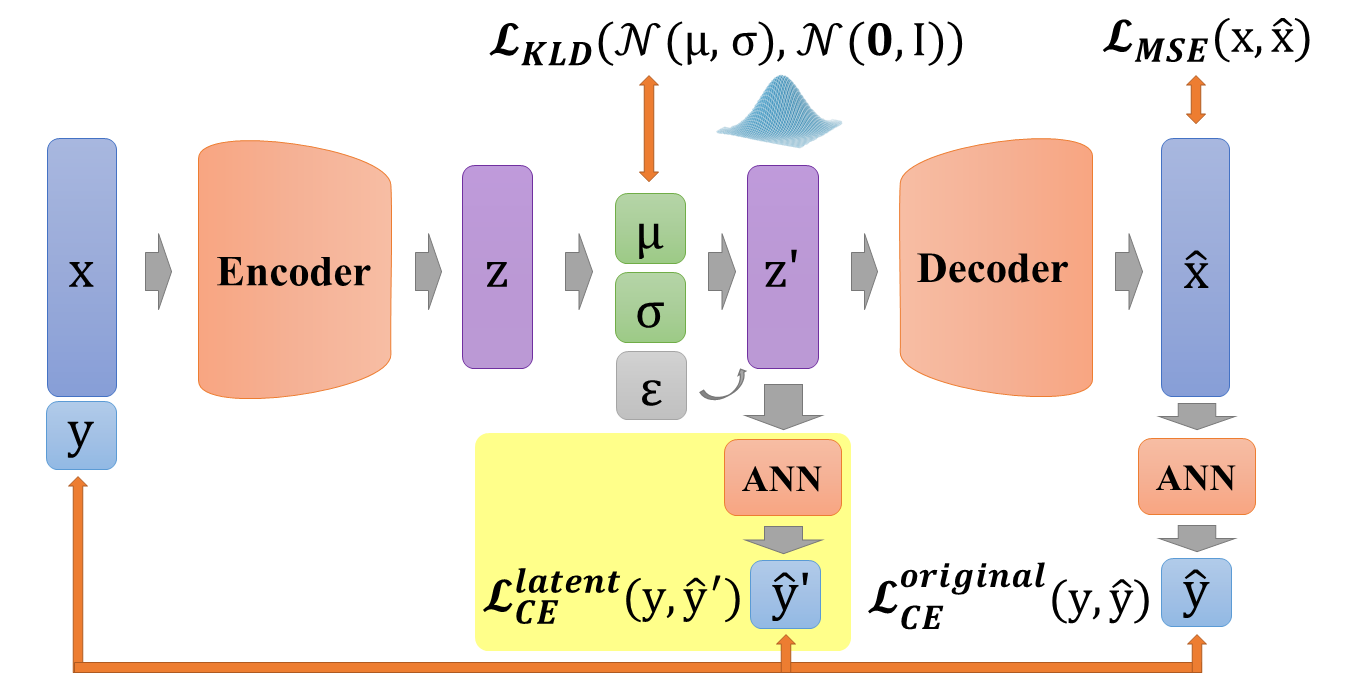}  \caption{Double Space \acrshort*{cvae} (\acrshort*{dscvae})}\label{fig:dscvae}
\end{figure}
Inspired by the latent discriminator from GET-3D~\cite{gao2022get3d}, we propose our \acrshort*{dscvae}. 
\acrshort*{dscvae} should be a more robust generative model to finely discriminate the difference between similar distributions of conditions. The constraint only at the original space enforces more information stored in the decoder according to~\cite{he2022masked}. In contrast, \acrshort*{dscvae} ensures the consistency of generated data by simultaneously assigning conditions to both the latent distribution and the original-space distribution, which is a crucial factor in the quality determination of generative models. 

In \acrshort*{dscvae}, we first adopt two classifiers at two spaces to classify the latent representation $\mathbf{z'}$ and the original space reconstruction $\mathbf{\hat{x}}$ respectively according to their common label. Double space conditions are implemented by adding another \acrshort*{ann} classifier for resampled variables $\mathbf{z'}$ (shown in Fig.~\ref{fig:dscvae}), of which classification error is still measured by \acrshort*{ce} loss as
\begin{equation}
\mathcal{L}_{CE}^{latent} = -(\mathbf{\hat{y}'} \log \mathcal{U}+(1-\mathbf{\hat{y}'}) \log (1-\mathcal{U}))
\end{equation}
where, latent-space predicted label is $\mathbf{\hat{y}'}$ and probability is $\mathcal{U}$. 

These two classifiers contribute to distinguishing different labels, more than judging the similarity of outputs in conventional generative models. Furthermore, our novel latent classifier promotes \acrshort*{dscvae} to learn a more informative latent space than generative models with only original-space conditions. Thus, \acrshort*{dscvae} should better solve the conditional inputs ("coalescence" or "non-coalescence" here) with considerable noise and similar distributions for different labels. 

The total loss function can be written as
\begin{align}
   \mathcal{L} = \mathcal{L}_{MSE}(\mathbf{x}, \hat{\mathbf{x}}) + \mathcal{L}_{KLD}(q(\mathbf{z|x,y}) \| p(\mathbf{z|x,y})) + \mathcal{L}_{CE}^{original} + \mathcal{L}_{CE}^{latent}
\end{align}
where $\mathcal{L}_{KLD}$ and $\mathcal{L}_{CE}^{latent}$  both regularise the latent distribution, making the latent space more interpretable. 

\section{Numerical Results and analysis}\label{NumericalResults}
In this section, we compare and analyze the performance of different generative and predictive methods. Ablation tests are performed to demonstrate the strength of the proposed \acrshort*{dscvae} model. \acrshort*{shap} value is used as a metric to interpret the model performance.
\subsection{Implementation details} 
In the generation phase, the balanced training set (shown in Table~\ref{tab:dataset}, named as Initial dataset hereafter) is employed in generative models as inputs and also used to inspect and validate these models. When training the generative neural networks, the parameters are set as follows: batch size is 73, the learning rate is $10^{-3}$, the optimizer is Adam, the scheduler is CosineAnnealingLR~\cite{loshchilov2016sgdr}, and the training epoch is 5000. For all generative methods, $\mathbf{z'}$ is resampled from the distribution, and added to a random four-dimensional Gaussian noise for regularizing purposes. The sum is put into the decoder to get the output $\mathbf{\hat{x}}$. The samples of different labels are generated separately. Totally, we generate 6570 balanced samples from each generative model. 

We then use the predictive models to evaluate the generated synthetic data. The training datasets are set as Table~\ref{tab:training}, and the validation set and test set are also shown in Table~\ref{tab:dataset}. 

\begin{table}[!htb]
\centering
\caption{Training Dataset}
\begin{tabular}{rccc}
\hline
\multicolumn{1}{l}{} & Initial & Synthetic & Total \\ \hline
\textbf{Initial dataset} & 438 & - & 438 \\
\textbf{\acrshort*{cvae} mixed dataset} & 438 & $438\times15$ & 7008 \\
\textbf{\acrshort*{dscvae} mixed dataset} & 438 & $438\times15$ & 7008 \\ \hline
\end{tabular}
\label{tab:training}
\end{table}

The algorithms for assessments are Random Forest (mentioned in section~\ref{mtd:rf}) and \acrlong*{xgb} (mentioned in section~\ref{mtd:xgb}). The key hyperparameters $\mathbf{n}_\textrm{estimators}$ and $\mathbf{d}_\textrm{max}$ are chosen from a grid search.
All the numerical experiments are implemented on the Google Colab platform. The \acrshort*{dl} generative models are performed on a single Tesla K80 GPU. 

\begin{figure}[h]
    \centering
    \subfloat[\acrshort*{mse}]{\includegraphics[width = 1.7 in]{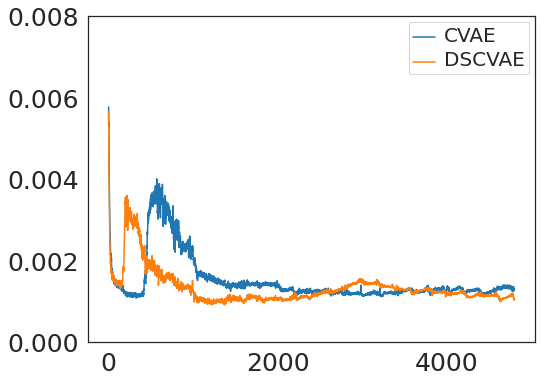} \label{fig:mse2}}
    \subfloat[\acrshort*{kld}]{\includegraphics[width = 1.7 in]{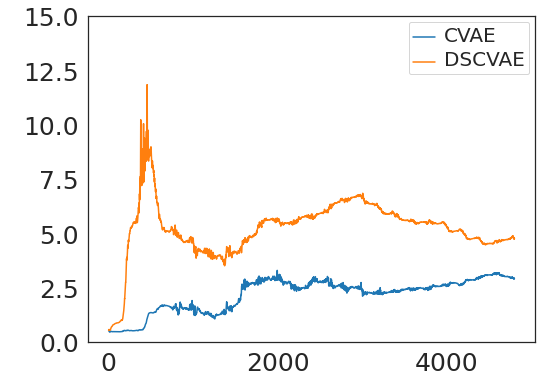} \label{fig:kld2}}\\
    \subfloat[CE of original space]{\includegraphics[width = 1.7 in]{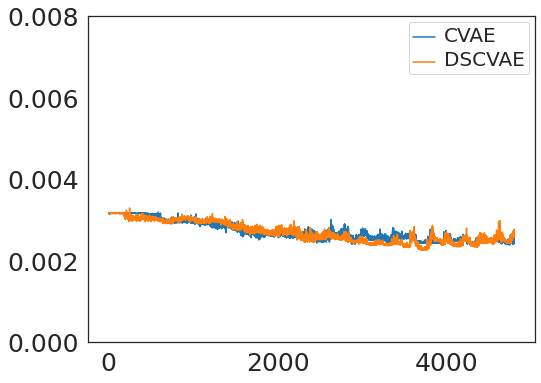} \label{fig:full2}}
    \subfloat[CE of latent space]{\includegraphics[width = 1.7 in]{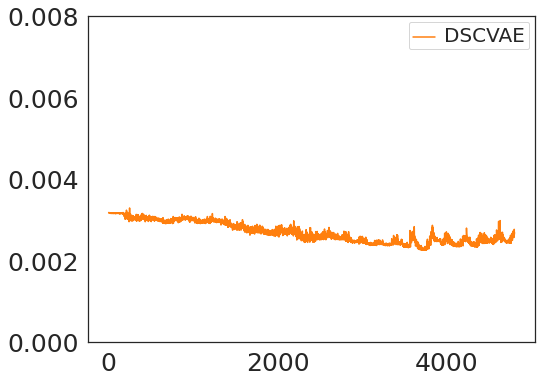} \label{fig:latent2}}
    \caption{Losses of the generative models}
    \label{fig:loss2}
\end{figure}

\subsection{Training loss of the generative models} 
The losses of \acrshort*{mse}, \acrshort*{kld} and classifiers both in the original space and the latent space are plotted in Fig.\ref{fig:loss2}. 
The \acrshort*{mse} loss and the \acrshort*{kld} loss are the basic loss functions used to build the VAE (Fig.\ref{fig:vae}). As shown in Fig.\ref{fig:mse2}, \acrshort*{mse} losses of both models tend to converge. Fig.\ref{fig:kld2} shows that the plots of \acrshort*{kld} losses vary significantly between \acrshort*{cvae} and \acrshort*{dscvae}.  The classifiers are used to discriminate the conditions (i.e., the label "coalescence" or "non-coalescence"). Fig.\ref{fig:full2} shows the classifiers of \acrshort*{cvae} and \acrshort*{dscvae} set for decoder outputs at the original space, while Fig.\ref{fig:latent2} shows the classifier of \acrshort*{dscvae} for the resampled variables in the latent space. The classifier losses  all decrease slightly before stabilization since the classifiers are used as regularizers in \acrshort*{cvae}s. 

The classifier of the latent space in \acrshort*{dscvae} has a direct impact on the construction of the latent distribution and indirectly affects the classification and reconstruction of the original space. Because the latent-space classifier affects the latent distribution, unsurprisingly we find that the \acrshort*{kld} loss of \acrshort*{dscvae} is more unstable than the one of \acrshort*{cvae}, as shown Fig.\ref{fig:kld2}. In addition, the double-space classifiers make the reconstruction training converge faster, as Fig.\ref{fig:mse2} shows that the \acrshort*{mse} loss of \acrshort*{dscvae} decreases faster than the \acrshort*{mse} loss of \acrshort*{cvae}. This may be due to the fact that the classifier of the latent space also guides the data reconstruction.

\begin{figure}[!h]
    \centering
    \subfloat[\acrshort*{rf}: Initial]{\includegraphics[width=1.7 in]{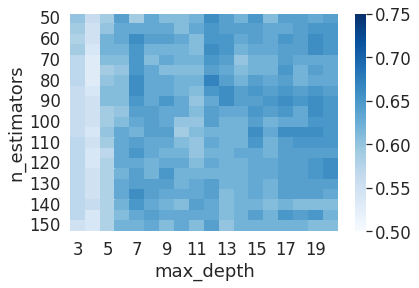} \label{fig:b_rf_heatmap}}
    \subfloat[\acrshort*{xgb}: Initial]{\includegraphics[width=1.7 in]{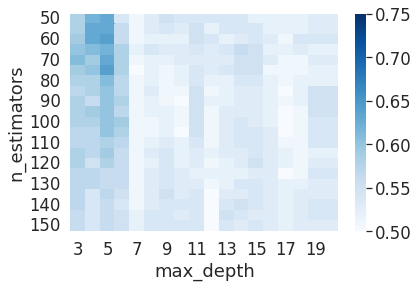} \label{fig:b_xgb_heatmap}}\\
    \subfloat[\acrshort*{rf}: \acrshort*{cvae}]{\includegraphics[width=1.7 in]{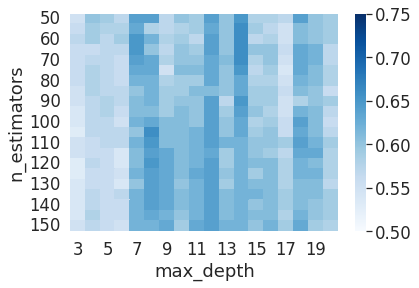} \label{fig:cvae_rf_heatmap}}
    \subfloat[\acrshort*{xgb}: \acrshort*{cvae}]{\includegraphics[width=1.7 in]{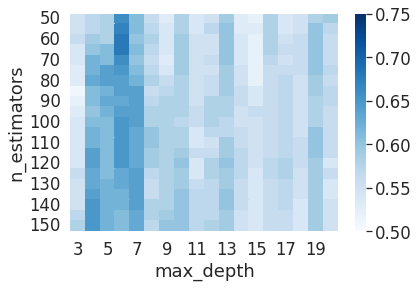} \label{fig:cvae_xgb_heatmap}}\\
    \subfloat[\acrshort*{rf}: \acrshort*{dscvae}]{\includegraphics[width=1.7 in]{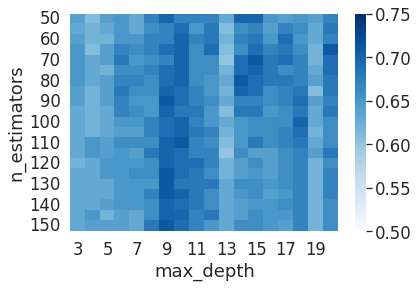} \label{fig:dscvae_rf_heatmap}}
    \subfloat[\acrshort*{xgb}: \acrshort*{dscvae}]{\includegraphics[width=1.7 in]{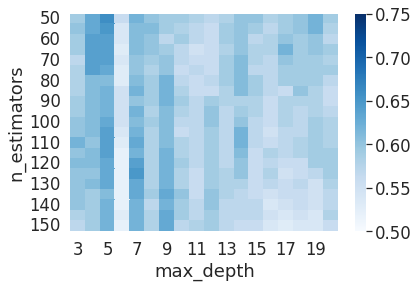} \label{fig:dscvae_xgb_heatmap}}
    \caption{Validation heatmaps for the tuning hyperparameters of the various generative models}
        \label{fig:heatmap}
\end{figure}

\subsection{Hyperparameter tuning}
The parameterization of \acrlong*{ml} algorithms impacts their generalization ability and  prediction accuracy. In this study, the balanced validation dataset is used to find the appropriate hyperparameters, namely $\mathbf{n}_\textrm{estimators}$ and $\mathbf{d}_\textrm{max}$. As shown in Fig.\ref{fig:heatmap}, the dark blue areas indicate high prediction accuracy in the validation set. For the \acrshort*{rf} predictors, the heatmap of \acrshort*{dscvae} (Fig.\ref{fig:dscvae_rf_heatmap}) has the largest dark areas, signifying higher prediction accuracy in this validation dataset. As for \acrlong*{xgb} predictors, \acrshort*{cvae} (Fig.\ref{fig:cvae_xgb_heatmap}) and \acrshort*{dscvae} (Fig.\ref{fig:dscvae_xgb_heatmap}) both lead to significantly more accurate predictions compared to solely using the original dataset(\ref{fig:b_xgb_heatmap}). Overall, the predictive models training on \acrshort*{dscvae} synthetic data show a more significant improvement in \acrshort*{rf} compared to XGboost in the validation dataset.
\begin{table*}[tb]
\centering
\caption{Validation results for hyperparameter tuning}
\resizebox{0.75\textwidth}{!}{%
\begin{tabular}{cccccccc}
\hline
\textbf{\begin{tabular}[c]{@{}c@{}}Evaluative\\ Method\end{tabular}} & \textbf{\begin{tabular}[c]{@{}c@{}}Generative\\ Method\end{tabular}} & \textbf{Accuracy ($\%$)} & \textbf{Precision ($\%$)} & \textbf{Recall ($\%$)} & \textbf{F1 score ($\%$)} & $\mathbf{n}_\textrm{estimators}$ & $\mathbf{d}_\textrm{max}$ \\ \hline
\multirow{4}{*}{\acrshort*{rf}} & None & 67.00 & 67.34 & 67.00 & 66.84 & 80 & 12 \\
 & \acrshort*{cvae} & 66.00 & 66.10 & 66.00 & 65.95 & 105 & 8 \\
 & \acrshort*{dscvae} & 71.00 & 71.01 & 71.00 & 71.00 & 125 & 9 \\ \hline
\multirow{4}{*}{\acrshort*{xgb}} & None & 64.00 & 64.09 & 64.00 & 63.94 & 60 & 5 \\
 & \acrshort*{cvae} & 68.00 & 68.12 & 68.00 & 67.95 & 60 & 6 \\
 & \acrshort*{dscvae} & 66.00 & 66.03 & 66.00 & 65.99 & 50 & 5 \\ \hline
\end{tabular}%
}
\label{tab:valid}
\end{table*}

\begin{table*}[!htb]
\centering
\caption{Test results using Initial or mixed datasets}
\resizebox{0.6\textwidth}{!}{%
\begin{tabular}{cccccc}
\hline
\textbf{\begin{tabular}[c]{@{}c@{}}Evaluative\\ Method\end{tabular}} & \textbf{\begin{tabular}[c]{@{}c@{}}Generative\\ Method\end{tabular}} & \textbf{Accuracy (\%)} & \textbf{Precision (\%)} & \textbf{Recall (\%)} & \textbf{F1 score (\%)} \\ \hline
\multirow{4}{*}{\acrshort*{rf}} & None & 58.50 & 58.57 & 58.50 & 58.42 \\
 & \acrshort*{cvae} & 63.00 & 63.13 & 63.00 & 62.91 \\
 & \acrshort*{cvae} (L) & 60.50 & 60.59 & 60.50 & 60.42 \\
 & \acrshort*{dscvae} & \textbf{66.00} & \textbf{66.42} & \textbf{66.00} & \textbf{65.78} \\ \hline
\multirow{4}{*}{\acrshort*{xgb}} & None & 58.00 & 58.01 & 58.00 & 57.98 \\
 & \acrshort*{cvae} & 63.00 & 63.01 & 63.00 & 63.00 \\
 & \acrshort*{cvae} (L) & 61.50 & 61.59 & 61.50 & 61.42 \\
 & \acrshort*{dscvae} & \textbf{66.50} & \textbf{66.58} & \textbf{66.50} & \textbf{66.46} \\ \hline
\end{tabular}%
}
\label{tab:acc}
\end{table*}
We fix $\mathbf{n}_\textrm{estimators}$ and $\mathbf{d}_\textrm{max}$ that perform the best on the validation dataset (according to Fig.\ref{fig:heatmap}) since the test dataset should not be used for tuning hyperparameters. These hyperparameters are used for predicting the test data. The exact values of these tuned hyperparameters are shown in Table~\ref{tab:valid}. 

\subsection{Predictive results}
Table~\ref{tab:acc} shows the prediction results based on different synthetic training datasets for \acrshort*{rf} and \acrlong*{xgb}. To further examine the necessity of training two classifiers jointly in \acrshort*{dscvae}, we set up another \acrshort*{cvae}(L) as an ablation experiment. The \acrshort*{cvae}(L) mixed dataset combines the Initial dataset and the synthetic data generated from the \acrshort*{cvae} with only one classifier in the latent space. It has the same hyperparameter tuning process on the predictive model as the other three sets. 

As displayed in table~\ref{tab:acc}, the predictors trained by \acrshort*{dscvae} mixed (synthetic and initial) dataset have the best test results in terms of all metrics investigated. In fact, all generated datasets improve the accuracy of predictive models. Compared to the predictive models only trained on the Initial dataset, \acrshort*{dscvae} substantially enhances the prediction accuracy by $7.5\%$ and $8.5\%$ for \acrshort*{rf} and \acrlong*{xgb} respectively. The predictive models trained by \acrshort*{dscvae} mixed dataset also have the best performance in precision, recall and F1 score. In addition, \acrshort*{dscvae} helps to reduce predictors over-fitting between validation results and test results. As shown in table~\ref{tab:acc}, \acrshort*{dscvae} synthetic data narrows the gap between validation accuracy and testing accuracy from $8.5\%$ to $5\%$ (\acrshort*{rf}) and from $6\%$ to $-0.5\%$ (\acrlong*{xgb}) compared to the Initial dataset. 

\begin{table}[!htb]
\centering
\caption{Confusion matrices for RF and XGBoost. Here, `TP' and `TN', and `FP' and `FN' correspond to true positive and true negative predictions, and false positive and false negative predictions, respectively, where ``positive" and ``negative", in turn, correspond to coalescence and non-coalescence events, respectively.}
\resizebox{0.48\textwidth}{!}{%
\begin{tabular}{cccccccccc}
\hline
\multirow{2}{*}{\textbf{\begin{tabular}[c]{@{}c@{}}Evaluative\\ Method\end{tabular}}} & \multirow{2}{*}{\textbf{\begin{tabular}[c]{@{}c@{}}Generative\\ Method\end{tabular}}} & \multicolumn{4}{c}{\textbf{Validation}} & \multicolumn{4}{c}{\textbf{Test}} \\
 &  & \textbf{TP} & \textbf{TN} & \textbf{FP} & \textbf{FN} & \textbf{TP} & \textbf{TN} & \textbf{FP} & \textbf{FN} \\ \hline
\multirow{4}{*}{\acrshort*{rf}} & None & 30 & 37 & 20 & 13 & 54 & 63 & 46 & 37 \\
 & \acrshort*{cvae} & 31 & 35 & 19 & 15 & 58 & 68 & 42 & 32 \\
 & \acrshort*{cvae} (L) & 30 & 38 & 20 & 12 & 56 & 65 & 44 & 35 \\
 & \acrshort*{dscvae} & 35 & 36 & 15 & 13 & \textbf{58} & \textbf{74} & 42 & 26 \\ \hline
\multirow{4}{*}{\acrshort*{xgb}} & None & 30 & 34 & 20 & 16 & 56 & 60 & 44 & 40 \\
 & \acrshort*{cvae} & 32 & 36 & 18 & 14 & 62 & 64 & 38 & 36 \\
 & \acrshort*{cvae} (L) & 28 & 38 & 22 & 12 & 57 & 66 & 43 & 34 \\
 & \acrshort*{dscvae} & 32 & 34 & 18 & 16 & \textbf{63} & \textbf{70} & 37 & 30 \\ \hline
\end{tabular}%
}
\label{tab:cm}
\end{table}

We consider "coalescence" as positive samples and "non-coalescence" as negative samples. Shown in Table~\ref{tab:cm} is the number of correct and incorrect predictions. The accurate predictions for both types increase compared to the predictive models trained on the Initial dataset. The true positive rates increase by $7.4\%$ and $12.5\%$ respectively, and the true negative rates increase by about $17\%$ for both \acrshort*{rf} and \acrlong*{xgb}. It is worth mentioning that the prediction accuracy of 66\% for DSCVAE might still present some limitations for the practical use of the predictive model. However, increasing the accuracy from 58\% to 66\% for a binary classification problem is still significant and it clearly demonstrates the strength of the proposed generative model. With this model, we predict the outcome, "coalescence" or "non-coalescence", of each forming doublet. Due to the dependence on the local process parameters, the probability of drop coalescence varies between 0 and 1, leaving some cases indeterminable. As a result, the predictability of the model cannot be very high. On the other hand, the reliable generation of a considerable amount of synthetic data will enable the prediction of coalescence probability for the prescribed set of data. Furthermore, a more accurate predictive model results in a more meaningful interpretability analysis as detailed in the following section.

\subsection{Intrepretability}
\begin{figure*}[!htb]
    \centering
    \subfloat[\acrshort*{rf}: Initial]{\includegraphics[width=1.15 in]{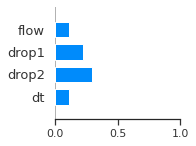} \label{fig:b_rf_shap_bar}}
    \subfloat[\acrshort*{rf}: \acrshort*{cvae} mixed]{\includegraphics[width=1.15 in]{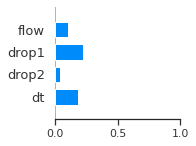} \label{fig:cvae_rf_shap_bar}}
    \subfloat[\acrshort*{rf}: \acrshort*{dscvae} mixed]{\includegraphics[width=1.15 in]{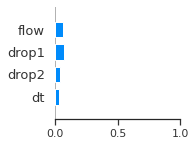} \label{fig:dscvae_rf_shap_bar}}
    \subfloat[\acrshort*{xgb}: Initial]{\includegraphics[width=1.15 in]{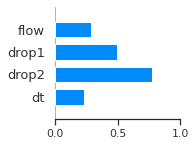} \label{fig:b_xgb_shap_bar}}
    \subfloat[\acrshort*{xgb}: \acrshort*{cvae} mixed]{\includegraphics[width=1.15 in]{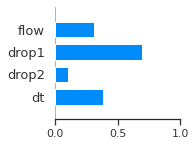} \label{fig:cvae_xgb_shap_bar}}
    \subfloat[\acrshort*{xgb}: \acrshort*{dscvae} mixed]{\includegraphics[width=1.15 in]{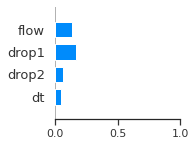} \label{fig:dscvae_xgb_shap_bar}}\\
    \subfloat[\acrshort*{rf}: Initial]{\includegraphics[width=1.15 in]{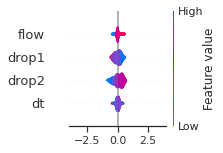} \label{fig:b_rf_shap_violin}}
    \subfloat[\acrshort*{rf}: \acrshort*{cvae} mixed]{\includegraphics[width=1.15 in]{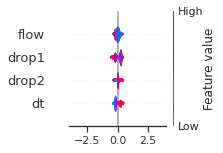} \label{fig:cvae_rf_shap_violin}} 
    \subfloat[\acrshort*{rf}: \acrshort*{dscvae} mixed]{\includegraphics[width=1.15 in]{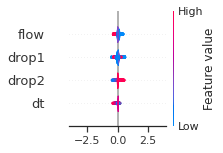} \label{fig:dscvae_rf_shap_violin}}
    \subfloat[\acrshort*{xgb}: Initial]{\includegraphics[width=1.15 in]{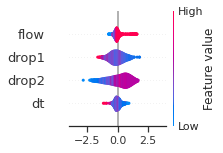} \label{fig:b_xgb_shap_violin}}
    \subfloat[\acrshort*{xgb}: \acrshort*{cvae} mixed]{\includegraphics[width=1.15 in]{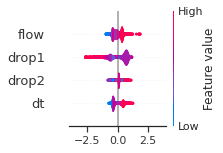} \label{fig:cvae_xgb_shap_violin}}
    \subfloat[\acrshort*{xgb}: \acrshort*{dscvae} mixed]{\includegraphics[width=1.15 in]{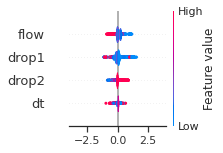} \label{fig:dscvae_xgb_shap_violin}}
    \caption{SHAP values for Training Dataset}
    \label{fig:shapvalue}
\end{figure*}
We use \acrshort*{shap} value here to reveal the contributions of samples in training datasets. The values can be analyzed both in global and local aspects. 
The bar plots (Fig.\ref{fig:shapvalue} (a-f)) show the mean of \acrshort*{shap} values of each feature. The models trained on the \acrshort*{dscvae} mixed (i.e., synthetic and initial) data have more similar \acrshort*{shap} values for different inputs. In Fig.\ref{fig:dscvae_rf_shap_bar} and Fig.\ref{fig:dscvae_xgb_shap_bar}, the \acrshort*{shap} values of all four features are relatively close compared to other approaches, indicating that the contributions of these features to the predictive models are similar. For example, the two features "drop1" and "drop2", in
Fig.\ref{fig:cvae_rf_shap_bar} and Fig.\ref{fig:cvae_xgb_shap_bar} show a large gap between them, which may be caused by overfitting in \acrshort*{cvae}. Whereas, as shown in Fig.\ref{fig:dscvae_rf_shap_bar},~\ref{fig:dscvae_xgb_shap_bar}, the \acrshort*{shap} values of "drop1" and "drop2" are more similar, and thus, more realistic. 

\begin{figure}[!htb]
  \centering
  \includegraphics[width= 2.5 in]{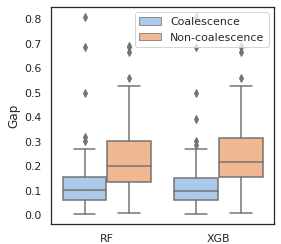}  \caption{"Gap" comparison according to predictive labels from \acrshort*{dscvae} enhanced predictors}\label{fig:shapexplain}
\end{figure}
In addition, scatter plots (Fig.\ref{fig:shapvalue} (g-l)) combine feature importances and feature effects. Each point on the scatter plot is a \acrshort*{shap} value of the feature in a training sample. The position on the x-axis is determined by the \acrshort*{shap} value which represents the feature contribution. The colours of the scatter points represent the feature values (red/blue for positive/negative contribution to "coalescence"). Most of the points in Fig.\ref{fig:dscvae_rf_shap_violin} and Fig.\ref{fig:dscvae_xgb_shap_violin} are clustered, indicating that their \acrshort*{shap} values are closer. This means that the individual samples in each feature contribute similarly to the model, which improves the model's robustness. Also, it can be seen from Fig.~\ref{fig:shapvalue}(i,l) that the contribution of "drop1" and "drop2" are reversed. That is, the higher size of "drop1" contributes negatively to the probability of "coalescence", while the higher size of "drop2" contributes positively to the probability of "coalescence". In fact, it can be noticed from Fig.~\ref{fig:dataset} that in general, the average size of "drop1" is larger than the one of "drop2". In addition, when the sizes of "drop1" and "drop2" are close, the drops have a higher probability of coalescence. To further confirm this trend, we plot in Fig.~\ref{fig:shapexplain} the impact of the drop size difference (i.e., $|\textbf{x}_\textrm{drop1}-\textbf{x}_\textrm{drop2}|$) on the model predictions of \acrshort*{rf} and \acrlong*{xgb}. 
It can be concluded that smaller differences in drop sizes can lead to a significantly higher probability of drop coalescence. In other words, to obtain a higher drop coalescence in experiments, one should minimize the drop size difference. This explains the opposite contribution of "drop1" size and "drop2" size in Fig.~\ref{fig:shapvalue} and also confirms the consistency of \acrshort*{dscvae} since both the \acrshort*{rf} and the \acrlong*{xgb} here are trained using the synthetic data. For the predictive models trained on the initial and \acrshort*{dscvae} datasets, it can be seen that the impact of $\textbf{x}_\textrm{dt}$ on model output is limited. It is worth mentioning that we consider here only the case when two drops form a doublet. For large $\textbf{x}_\textrm{dt}$, drops can proceed as singlets, without forming a doublet and therefore are not included in the consideration.

In summary, \acrshort*{dscvae} mixed data outperforms the Initial dataset and \acrshort*{cvae} mixed dataset in training predictors in classification tasks. It not only reduces the overfitting gap between validation accuracy and test accuracy, but also increases overall predictive performance. By analyzing \acrlong*{xgb} values of each predictive model, we find that \acrshort*{dscvae} mixed dataset contributes to features' homogeneity, allowing features to be treated equally. 

\section{Conclusions}
In this study,  we propose a novel generative model named the \acrfull*{dscvae} for generating synthetic tabular data. Compared to traditional generative models, the proposed model utilizes two classifiers in the latent and original spaces to regularize the data generation process. The \acrshort*{dscvae} model is applied to generate synthetic data for training predictive models of drop coalescence in a microfluidics device. Numerical results show that the predictive models, namely \acrshort*{rf} and \acrlong*{xgb}, achieve substantially more accurate and robust predictions than those trained on purely experimental data in terms of several metrics. With the help of \acrshort*{dscvae} synthetic data, both \acrshort*{rf} and \acrlong*{xgb} manage to obtain a prediction accuracy of about $66\%$, which is about $8\%$ higher than using only the experimental data. It is worth mentioning that, from a prediction perspective, this is an extremely challenging task due to, firstly the imbalanced dataset, and more importantly, the similar input distributions of "coalescence" and "non-coalescence" samples. 
These predictive models are valuable for optimizing experimental design, and insightful analysis is provided through the computation of \acrfull*{shap} values.  The results show that reducing the gap between two drop sizes increases the probability of coalescence. Moreover, ablation tests demonstrate the strength of \acrshort*{dscvae} compared to non-conditional \acrshort*{vae} and standard \acrshort*{cvae} (with constraints either in the latent or the original space). 

This study offers a paradigm for addressing classification problems with limited and imbalanced tabular data. It also highlights the potential of data-driven methods for predicting microfluidic drop coalescence. The proposed methodology can be applied to other materials and devices. Future work may include further improving the interpretability and robustness of \acrshort*{dscvae}, for instance, by imposing physical or chemical constraints. The future work will also aim at extending the model parametric space to include viscosities of continuous and dispersed phases, interfacial tension, dynamic surfactant effects, and the presence of surfactants in different phases.

\section*{Code and data availability}

The computational part of this study is performed using Python language. The code is available at: \url{https://github.com/DL-WG/dscvae-for-drop-coalescence}.

Experimental data is available upon reasonable request to Dr Nina Kovalchuk (n.kovalchuk@bham.ac.uk)

\section*{Acknowledgement}\label{Acknowledgement}

This research is funded by the EP/T000414/1 PREdictive Modelling with
Quantification of UncERtainty for MultiphasE Systems (PREMIERE). This work is partially supported by the Leverhulme Centre for Wildfires, Environment and Society through the Leverhulme Trust, grant number RC-2018-023.

\printglossary[type=\acronymtype, nonumberlist]



\balance

\renewcommand\refname{References}

\bibliography{references0} 

\begin{thebibliography}{10}
\providecommand{\url}[1]{#1}
\csname url@samestyle\endcsname
\providecommand{\newblock}{\relax}
\providecommand{\bibinfo}[2]{#2}
\providecommand{\BIBentrySTDinterwordspacing}{\spaceskip=0pt\relax}
\providecommand{\BIBentryALTinterwordstretchfactor}{4}
\providecommand{\BIBentryALTinterwordspacing}{\spaceskip=\fontdimen2\font plus
\BIBentryALTinterwordstretchfactor\fontdimen3\font minus
  \fontdimen4\font\relax}
\providecommand{\BIBforeignlanguage}[2]{{%
\expandafter\ifx\csname l@#1\endcsname\relax
\typeout{** WARNING: IEEEtran.bst: No hyphenation pattern has been}%
\typeout{** loaded for the language `#1'. Using the pattern for}%
\typeout{** the default language instead.}%
\else
\language=\csname l@#1\endcsname
\fi
#2}}
\providecommand{\BIBdecl}{\relax}
\BIBdecl

\bibitem{krebs2012microfluidic}
T.~Krebs, K.~Schroen, and R.~Boom, ``A microfluidic method to study
  demulsification kinetics,'' \emph{Lab on a Chip}, vol.~12, no.~6, pp.
  1060--1070, 2012.

\bibitem{liu2016droplet}
Z.~Liu, X.~Wang, R.~Cao, and Y.~Pang, ``Droplet coalescence at microchannel
  intersection chambers with different shapes,'' \emph{Soft Matter}, vol.~12,
  no.~26, pp. 5797--5807, 2016.

\bibitem{shenoy2016stokes}
A.~Shenoy, C.~V. Rao, and C.~M. Schroeder, ``Stokes trap for multiplexed
  particle manipulation and assembly using fluidics,'' \emph{Proceedings of the
  National Academy of Sciences}, vol. 113, no.~15, pp. 3976--3981, 2016.

\bibitem{niu2008pillar}
X.~Niu, S.~Gulati, J.~B. Edel, and A.~J. demello, ``Pillar-induced droplet
  merging in microfluidic circuits,'' \emph{Lab on a Chip}, vol.~8, no.~11, pp.
  1837--1841, 2008.

\bibitem{keith2021combining}
J.~A. Keith, V.~Vassilev-Galindo, B.~Cheng, S.~Chmiela, M.~Gastegger, K.-R.
  Mu\texttt{"u}ller, and A.~Tkatchenko, ``Combining machine learning and
  computational chemistry for predictive insights into chemical systems,''
  \emph{Chemical reviews}, vol. 121, no.~16, pp. 9816--9872, 2021.

\bibitem{dral2020quantum}
P.~O. Dral, ``Quantum chemistry in the age of machine learning,'' \emph{The
  journal of physical chemistry letters}, vol.~11, no.~6, pp. 2336--2347, 2020.

\bibitem{meuwly2021machine}
M.~Meuwly, ``Machine learning for chemical reactions,'' \emph{Chemical
  Reviews}, vol. 121, no.~16, pp. 10\,218--10\,239, 2021.

\bibitem{wikramanayake2020statistical}
E.~Wikramanayake and V.~Bahadur, ``Statistical modeling of
  electrowetting-induced droplet coalescence for condensation applications,''
  \emph{Colloids and Surfaces A: Physicochemical and Engineering Aspects}, vol.
  599, p. 124874, 2020.

\bibitem{rodriguez2022parameterization}
C.~F. Rodr{\'\i}guez~Gen{\'o} and L.~Alfonso, ``Parameterization of the
  collision--coalescence process using series of basis functions: Colnetv1. 0.0
  model development using a machine learning approach,'' \emph{Geoscientific
  Model Development}, vol.~15, no.~2, pp. 493--507, 2022.

\bibitem{zhuang2022ensemble}
Y.~Zhuang, S.~Cheng, N.~Kovalchuk, M.~Simmons, O.~K. Matar, Y.-K. Guo, and
  R.~Arcucci, ``Ensemble latent assimilation with deep learning surrogate
  model: application to drop interaction in a microfluidics device,'' \emph{Lab
  on a Chip}, vol.~22, no.~17, pp. 3187--3202, 2022.

\bibitem{he2009learning}
H.~He and E.~A. Garcia, ``Learning from imbalanced data,'' \emph{IEEE
  Transactions on knowledge and data engineering}, vol.~21, no.~9, pp.
  1263--1284, 2009.

\bibitem{mitra2020multi}
S.~Mitra, S.~Saha, and M.~Hasanuzzaman, ``A multi-view deep neural network
  model for chemical-disease relation extraction from imbalanced datasets,''
  \emph{IEEE Journal of Biomedical and Health Informatics}, vol.~24, no.~11,
  pp. 3315--3325, 2020.

\bibitem{thakkar2018liver}
S.~Thakkar, M.~Chen, H.~Fang, Z.~Liu, R.~Roberts, and W.~Tong, ``The liver
  toxicity knowledge base (lktb) and drug-induced liver injury (dili)
  classification for assessment of human liver injury,'' \emph{Expert review of
  gastroenterology \& hepatology}, vol.~12, no.~1, pp. 31--38, 2018.

\bibitem{su2015improving}
C.~Su, S.~Ju, Y.~Liu, and Z.~Yu, ``Improving random forest and rotation forest
  for highly imbalanced datasets,'' \emph{Intelligent Data Analysis}, vol.~19,
  no.~6, pp. 1409--1432, 2015.

\bibitem{blaszczynski2015neighbourhood}
J.~B{\l}aszczy{\'n}ski and J.~Stefanowski, ``Neighbourhood sampling in bagging
  for imbalanced data,'' \emph{Neurocomputing}, vol. 150, pp. 529--542, 2015.

\bibitem{bader2018biased}
M.~Bader-El-Den, E.~Teitei, and T.~Perry, ``Biased random forest for dealing
  with the class imbalance problem,'' \emph{IEEE transactions on neural
  networks and learning systems}, vol.~30, no.~7, pp. 2163--2172, 2018.

\bibitem{ferreira2019adaptive}
L.~E.~B. Ferreira, H.~M. Gomes, A.~Bifet, and L.~S. Oliveira, ``Adaptive random
  forests with resampling for imbalanced data streams,'' in \emph{2019
  International Joint Conference on Neural Networks (IJCNN)}.\hskip 1em plus
  0.5em minus 0.4em\relax IEEE, 2019, pp. 1--6.

\bibitem{batista2004study}
G.~E. Batista, R.~C. Prati, and M.~C. Monard, ``A study of the behavior of
  several methods for balancing machine learning training data,'' \emph{ACM
  SIGKDD explorations newsletter}, vol.~6, no.~1, pp. 20--29, 2004.

\bibitem{wei2019eda}
J.~Wei and K.~Zou, ``Eda: Easy data augmentation techniques for boosting
  performance on text classification tasks,'' \emph{arXiv preprint
  arXiv:1901.11196}, pp. 6383--–6389, 2019.

\bibitem{goodfellow2020generative}
I.~Goodfellow, J.~Pouget-Abadie, M.~Mirza, B.~Xu, D.~Warde-Farley, S.~Ozair,
  A.~Courville, and Y.~Bengio, ``Generative adversarial networks,''
  \emph{Communications of the ACM}, vol.~63, no.~11, pp. 139--144, 2020.

\bibitem{moon2020conditional}
J.~Moon, S.~Jung, S.~Park, and E.~Hwang, ``Conditional tabular gan-based
  two-stage data generation scheme for short-term load forecasting,''
  \emph{IEEE Access}, vol.~8, pp. 205\,327--205\,339, 2020.

\bibitem{burks2019data}
R.~Burks, K.~A. Islam, Y.~Lu, and J.~Li, ``Data augmentation with generative
  models for improved malware detection: A comparative study,'' in \emph{2019
  IEEE 10th Annual Ubiquitous Computing, Electronics \& Mobile Communication
  Conference (UEMCON)}.\hskip 1em plus 0.5em minus 0.4em\relax IEEE, 2019, pp.
  0660--0665.

\bibitem{kingma2014auto}
D.~P. Kingma and M.~Welling, ``Auto-encoding variational bayes,'' \emph{stat},
  vol. 1050, p.~1, 2014.

\bibitem{sohn2015learning}
\BIBentryALTinterwordspacing
K.~Sohn, H.~Lee, and X.~Yan, ``Learning structured output representation using
  deep conditional generative models,'' in \emph{Advances in Neural Information
  Processing Systems}, C.~Cortes, N.~Lawrence, D.~Lee, M.~Sugiyama, and
  R.~Garnett, Eds., vol.~28.\hskip 1em plus 0.5em minus 0.4em\relax Curran
  Associates, Inc., 2015. [Online]. Available:
  \url{https://proceedings.neurips.cc/paper/2015/file/8d55a249e6baa5c06772297520da2051-Paper.pdf}
\BIBentrySTDinterwordspacing

\bibitem{huang2022ada}
K.~Huang and X.~Wang, ``Ada-incvae: Improved data generation using variational
  autoencoder for imbalanced classification,'' \emph{Applied Intelligence},
  vol.~52, no.~3, pp. 2838--2853, 2022.

\bibitem{chen2021trajvae}
X.~Chen, J.~Xu, R.~Zhou, W.~Chen, J.~Fang, and C.~Liu, ``Trajvae: A variational
  autoencoder model for trajectory generation,'' \emph{Neurocomputing}, vol.
  428, pp. 332--339, 2021.

\bibitem{yang2019improving}
Y.~Yang, K.~Zheng, C.~Wu, and Y.~Yang, ``Improving the classification
  effectiveness of intrusion detection by using improved conditional
  variational autoencoder and deep neural network,'' \emph{Sensors}, vol.~19,
  no.~11, p. 2528, 2019.

\bibitem{he2022masked}
K.~He, X.~Chen, S.~Xie, Y.~Li, P.~Doll{\'a}r, and R.~Girshick, ``Masked
  autoencoders are scalable vision learners,'' in \emph{Proceedings of the
  IEEE/CVF Conference on Computer Vision and Pattern Recognition}, 2022, pp.
  16\,000--16\,009.

\bibitem{chagot2022surfactant}
L.~Chagot, C.~Quilodr{\'a}n-Casas, M.~Kalli, N.~M. Kovalchuk, M.~J. Simmons,
  O.~K. Matar, R.~Arcucci, and P.~Angeli, ``Surfactant-laden droplet size
  prediction in a flow-focusing microchannel: a data-driven approach,''
  \emph{Lab on a Chip}, vol.~22, no.~20, pp. 3848--3859, 2022.

\bibitem{gelada2019deepmdp}
C.~Gelada, S.~Kumar, J.~Buckman, O.~Nachum, and M.~G. Bellemare, ``Deepmdp:
  Learning continuous latent space models for representation learning,'' in
  \emph{International Conference on Machine Learning}.\hskip 1em plus 0.5em
  minus 0.4em\relax PMLR, 2019, pp. 2170--2179.

\bibitem{bojanowski2018optimizing}
P.~Bojanowski, A.~Joulin, D.~Lopez-Pas, and A.~Szlam, ``Optimizing the latent
  space of generative networks,'' in \emph{International Conference on Machine
  Learning}.\hskip 1em plus 0.5em minus 0.4em\relax PMLR, 2018, pp. 600--609.

\bibitem{gao2022get3d}
J.~Gao, T.~Shen, Z.~Wang, W.~Chen, K.~Yin, D.~Li, O.~Litany, Z.~Gojcic, and
  S.~Fidler, ``Get3d: A generative model of high quality 3d textured shapes
  learned from images,'' \emph{arXiv preprint arXiv:2209.11163}, 2022.

\bibitem{kim2008soft}
P.~Kim, K.~W. Kwon, M.~C. Park, S.~H. Lee, S.~M. Kim, and K.~Y. Suh, ``Soft
  lithography for microfluidics: a review,'' \emph{BIOCHIP JOURNAL}, vol.~2,
  no.~1, pp. 1--11, 2008.

\bibitem{kovalchuk2019drop}
N.~M. Kovalchuk, M.~Sagisaka, K.~Steponavicius, D.~Vigolo, and M.~J. Simmons,
  ``Drop formation in microfluidic cross-junction: jetting to dripping to
  jetting transition,'' \emph{Microfluidics and Nanofluidics}, vol.~23, no.~8,
  pp. 1--14, 2019.

\bibitem{schneider2012nih}
C.~A. Schneider, W.~S. Rasband, and K.~W. Eliceiri, ``Nih image to imagej: 25
  years of image analysis,'' \emph{Nature methods}, vol.~9, no.~7, pp.
  671--675, 2012.

\bibitem{buzzaccaro2013ghost}
S.~Buzzaccaro, E.~Secchi, and R.~Piazza, ``Ghost particle velocimetry: accurate
  3d flow visualization using standard lab equipment,'' \emph{Physical review
  letters}, vol. 111, no.~4, p. 048101, 2013.

\bibitem{kovalchuk2018study}
N.~Kovalchuk, J.~Chowdhury, Z.~Schofield, D.~Vigolo, and M.~Simmons, ``Study of
  drop coalescence and mixing in microchannel using ghost particle
  velocimetry,'' \emph{Chemical Engineering Research and Design}, vol. 132, pp.
  881--889, 2018.

\bibitem{yi2020efficient}
H.~Yi, C.~Zhu, T.~Fu, and Y.~Ma, ``Efficient coalescence of microdroplet in the
  cross-focused microchannel with symmetrical chamber,'' \emph{Journal of the
  Taiwan Institute of Chemical Engineers}, vol. 112, pp. 52--59, 2020.

\bibitem{grinsztajn2022tree}
L.~Grinsztajn, E.~Oyallon, and G.~Varoquaux, ``Why do tree-based models still
  outperform deep learning on tabular data?'' \emph{arXiv preprint
  arXiv:2207.08815}, 2022.

\bibitem{patel2018study}
H.~H. Patel and P.~Prajapati, ``Study and analysis of decision tree based
  classification algorithms,'' \emph{International Journal of Computer Sciences
  and Engineering}, vol.~6, no.~10, pp. 74--78, 2018.

\bibitem{gong2022efficient}
H.~Gong, S.~Cheng, Z.~Chen, Q.~Li, C.~Quilodr{\'a}n-Casas, D.~Xiao, and
  R.~Arcucci, ``An efficient digital twin based on machine learning svd
  autoencoder and generalised latent assimilation for nuclear reactor
  physics,'' \emph{Annals of Nuclear Energy}, vol. 179, p. 109431, 2022.

\bibitem{pal2003assessment}
M.~Pal and P.~M. Mather, ``An assessment of the effectiveness of decision tree
  methods for land cover classification,'' \emph{Remote sensing of
  environment}, vol.~86, no.~4, pp. 554--565, 2003.

\bibitem{turney1995bias}
P.~Turney, ``Bias and the quantification of stability,'' \emph{Machine
  Learning}, vol.~20, no.~1, pp. 23--33, 1995.

\bibitem{ho1995random}
T.~K. Ho, ``Random decision forests,'' in \emph{Proceedings of 3rd
  international conference on document analysis and recognition}, vol.~1.\hskip
  1em plus 0.5em minus 0.4em\relax IEEE, 1995, pp. 278--282.

\bibitem{breiman1996bagging}
L.~Breiman, ``Bagging predictors,'' \emph{Machine learning}, vol.~24, no.~2,
  pp. 123--140, 1996.

\bibitem{cheng2022parameter}
S.~Cheng, Y.~Jin, S.~P. Harrison, C.~Quilodr{\'a}n-Casas, I.~C. Prentice, Y.-K.
  Guo, and R.~Arcucci, ``Parameter flexible wildfire prediction using machine
  learning techniques: Forward and inverse modelling,'' \emph{Remote Sensing},
  vol.~14, no.~13, p. 3228, 2022.

\bibitem{breiman2001random}
L.~Breiman, ``Random forests,'' \emph{Machine learning}, vol.~45, no.~1, pp.
  5--32, 2001.

\bibitem{friedman2001greedy}
J.~H. Friedman, ``Greedy function approximation: a gradient boosting machine,''
  \emph{Annals of statistics}, pp. 1189--1232, 2001.

\bibitem{chen2015xgboost}
T.~Chen, T.~He, M.~Benesty, V.~Khotilovich, Y.~Tang, H.~Cho, K.~Chen
  \emph{et~al.}, ``Xgboost: extreme gradient boosting,'' \emph{R package
  version 0.4-2}, vol.~1, no.~4, pp. 1--4, 2015.

\bibitem{kuhn1953contributions}
H.~W. Kuhn and A.~W. Tucker, \emph{Contributions to the Theory of Games}.\hskip
  1em plus 0.5em minus 0.4em\relax Princeton University Press, 1953, no.~28.

\bibitem{lundberg2020local2global}
S.~M. Lundberg, G.~Erion, H.~Chen, A.~DeGrave, J.~M. Prutkin, B.~Nair, R.~Katz,
  J.~Himmelfarb, N.~Bansal, and S.-I. Lee, ``From local explanations to global
  understanding with explainable ai for trees,'' \emph{Nature Machine
  Intelligence}, vol.~2, no.~1, pp. 2522--5839, 2020.

\bibitem{kramer1991nonlinear}
M.~A. Kramer, ``Nonlinear principal component analysis using autoassociative
  neural networks,'' \emph{AIChE journal}, vol.~37, no.~2, pp. 233--243, 1991.

\bibitem{cheng2023generalised}
S.~Cheng, J.~Chen, C.~Anastasiou, P.~Angeli, O.~K. Matar, Y.-K. Guo, C.~C.
  Pain, and R.~Arcucci, ``Generalised latent assimilation in heterogeneous
  reduced spaces with machine learning surrogate models,'' \emph{Journal of
  Scientific Computing}, vol.~94, no.~1, pp. 1--37, 2023.

\bibitem{cheng2022data}
S.~Cheng, I.~C. Prentice, Y.~Huang, Y.~Jin, Y.-K. Guo, and R.~Arcucci,
  ``Data-driven surrogate model with latent data assimilation: Application to
  wildfire forecasting,'' \emph{Journal of Computational Physics}, p. 111302,
  2022.

\bibitem{yousefi2017autoencoder}
M.~Yousefi-Azar, V.~Varadharajan, L.~Hamey, and U.~Tupakula,
  ``Autoencoder-based feature learning for cyber security applications,'' in
  \emph{2017 International joint conference on neural networks (IJCNN)}.\hskip
  1em plus 0.5em minus 0.4em\relax IEEE, 2017, pp. 3854--3861.

\bibitem{esmaeili2019structured}
B.~Esmaeili, H.~Wu, S.~Jain, A.~Bozkurt, N.~Siddharth, B.~Paige, D.~H. Brooks,
  J.~Dy, and J.-W. Meent, ``Structured disentangled representations,'' in
  \emph{The 22nd International Conference on Artificial Intelligence and
  Statistics}.\hskip 1em plus 0.5em minus 0.4em\relax PMLR, 2019, pp.
  2525--2534.

\bibitem{zhao2017learning}
T.~Zhao, R.~Zhao, and M.~Eskenazi, ``Learning discourse-level diversity for
  neural dialog models using conditional variational autoencoders,'' in
  \emph{Proceedings of the 55th Annual Meeting of the Association for
  Computational Linguistics (Volume 1: Long Papers)}, 2017, pp. 654--664.

\bibitem{loshchilov2016sgdr}
I.~Loshchilov and F.~Hutter, ``Sgdr: Stochastic gradient descent with warm
  restarts,'' \emph{arXiv e-prints}, pp. arXiv--1608, 2016.

\end{thebibliography}
\bibliographystyle{IEEEtran}

\end{document}